\documentclass[
reprint,
superscriptaddress,
amsmath,
amssymb,
aps,
prb,
floatfix,
]{revtex4-2}
\usepackage{graphicx}
\usepackage{hyperref}
\usepackage{xcolor}
\usepackage{multirow}

\begin{document}

\title{Formation of L1\textsubscript{2}-ordered $\gamma'$-Ni\textsubscript{3}Al precipitates in ternary Cu--Ni--Al alloys modelled using an \textit{ab initio} concentration wave theory and atomistic simulations}

\author{Christopher D. Woodgate}
\email{Christopher.Woodgate@bristol.ac.uk}
\affiliation{H.H. Wills Physics Laboratory, University of Bristol, Royal Fort, Bristol, BS8 1TL, United Kingdom}
\affiliation{Department of Physics, University of Warwick, Coventry, CV4 7AL, United Kingdom}
\author{Hubert J. Naguszewski}
\affiliation{Department of Physics, University of Warwick, Coventry, CV4 7AL, United Kingdom}
\author{Samuel L. Deacon}
\thanks{These authors contributed equally to this work.}
\affiliation{Department of Physics, University of Warwick, Coventry, CV4 7AL, United Kingdom}
\author{Mathys P. Potel}
\thanks{These authors contributed equally to this work.}
\affiliation{Department of Physics, University of Warwick, Coventry, CV4 7AL, United Kingdom}
\author{Ján Minár}
\affiliation{New Technologies Research Center, University of West Bohemia in Pilsen, Pilsen, Czech Republic}
\author{David Quigley}
\affiliation{Department of Physics, University of Warwick, Coventry, CV4 7AL, United Kingdom}
\author{Julie B. Staunton}
\email{J.B.Staunton@warwick.ac.uk}
\affiliation{Department of Physics, University of Warwick, Coventry, CV4 7AL, United Kingdom}

\begin{abstract}
Precipitation-strengthened Cu--Ni--Al alloys are of interest for technological applications because coherent, L1\textsubscript{2}-ordered $\gamma'$-Ni\textsubscript{3}Al precipitates can confer high mechanical strength while allowing the material to retain many of the good transport properties characteristic of elemental Cu.
In this work, we study the thermodynamics and phase stability of the pseudobinary $\textrm{Cu}_x (\textrm{Ni}_{3/4} \textrm{Al}_{1/4})_{1-x}$ system, $0 \leq x \leq 1$. 
We use a computational modelling framework combining first-principles electronic structure calculations with a concentration wave analysis from which atom-atom effective pair interactions are extracted for use in atomistic Monte Carlo simulations.
Our modelling reveals three distinct, composition-dependent regimes of phase behaviour, in qualitative agreement with the experimentally determined phase diagram.
At low Cu content, Cu is soluble in the L1\textsubscript{2}-ordered Ni\textsubscript{3}Al phase, with a single identifiable phase transition corresponding to chemical ordering between Ni and Al.
At intermediate compositions, this high-temperature ordering is followed at lower temperatures by phase separation of Cu and L1\textsubscript{2}-ordered Ni\textsubscript{3}Al.
Finally, at high Cu content, L1\textsubscript{2}-ordered Ni\textsubscript{3}Al precipitates directly from the solid solution, with no clearly identifiable secondary transition.
We  relate these phase transformations to features of the underlying electronic structures of the considered alloys.
Overall, this work demonstrates a computationally efficient workflow capturing both chemical ordering and coherent precipitation in multicomponent substitutional alloys, with relevance to the study of phenomena such as precipitation strengthening.
\end{abstract}

\date{July 29, 2026}

\maketitle

\section{Introduction}
\label{sec:introduction}

The alloying of metallic elements facilitates control of materials' structure at the atomic scale, leading to fine-tuning of a range of desirable properties for specific applications~\cite{campbell_elements_2008, callister_materials_2020}. Often it is of particular importance to understand how atoms in a given alloy arrange themselves when in thermal equilibrium at ambient pressure, \textit{i.e.}\ to determine the alloy's phase diagram. Depending on the chemical composition and the targeted application, advantageous properties may be conferred when the elements in an alloy mix freely (as in a solid solution), when the elements form some chemically ordered structure (as in an intermetallic compound), or when one or more elements in the composition cluster together (phase separation) and the alloy develops a chemically inhomogeneous microstructure at the nanoscale.

\begin{figure}[t]
    \centering
    \includegraphics[width=\linewidth]{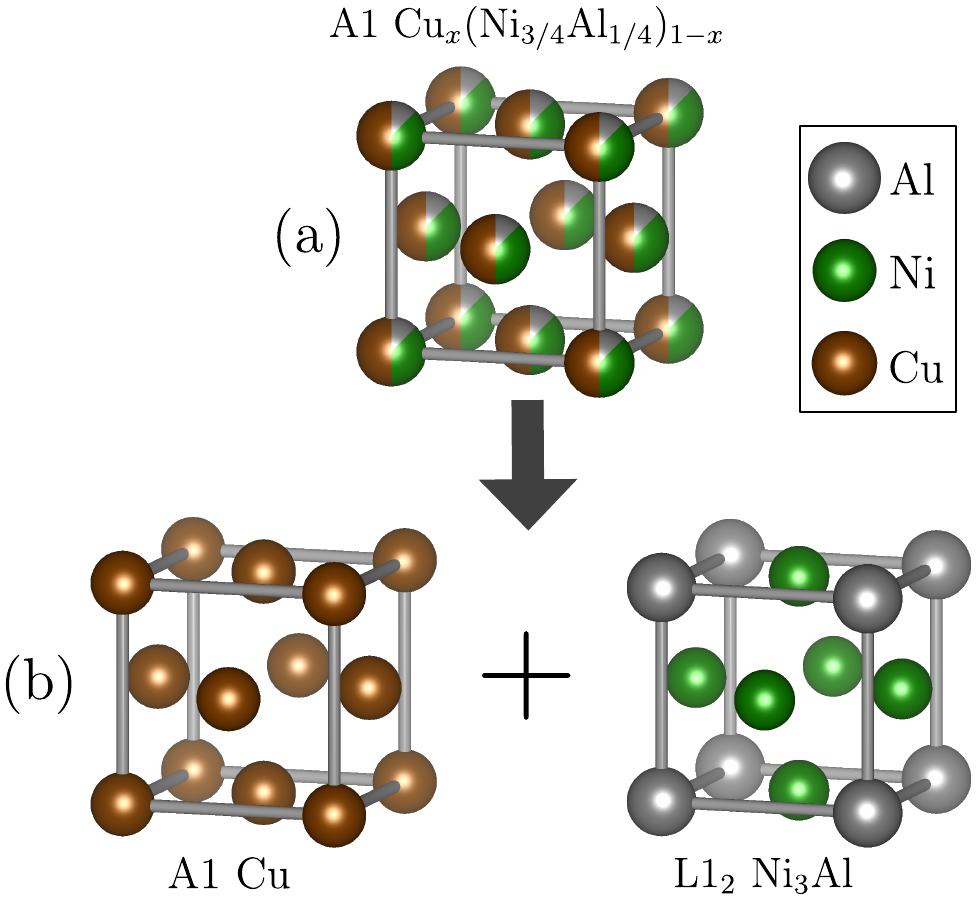}
    \caption{Conceptual illustration of (a) a disordered fcc (A1) Cu--Ni--Al alloy decomposing into (b) a mixture of fcc (A1) Cu (left) and L1\textsubscript{2}-ordered Ni\textsubscript{3}Al (right). This metallurgical reaction is the focus of the present study. Partially-coloured spheres indicate lattice sites where elements mix freely. Images generated using \textsc{VESTA}~\cite{momma_vesta_2011}.}
    \label{fig:cunial_schematic}
\end{figure}

A well-studied system in which both chemical ordering and phase separation occur is provided by ternary Cu--Ni--Al alloys~\cite{tsuda_improvement_1986, cho_precipitation_1991, grylls_strengthening_1996, sierpinski_phase_1999, wang_phase_2003, christofidou_effect_2017, li_comparative_2019, ferguson_continuous_2020, li_cuboidal_2021, li_strong_2021, semboshi_age-induced_2021, zhang_effect_2022, li_resistivity-temperature_2025, cheng_composition_2025} and---in particular---the $\textrm{Cu}_x(\textrm{Ni}_{3/4}\textrm{Al}_{1/4})_{1-x}$ pseudobinary system~\cite{semboshi_phase_2022}. Here, `pseudobinary' refers to a single-parameter path in composition space along which the Ni:Al ratio is held fixed at 3:1, while the fraction of Cu is controlled via the parameter $x$. Elemental Cu has exceptional electrical and thermal conductivity but is mechanically soft, with a low yield strength that limits its use in load-bearing applications. The binary Ni\textsubscript{3}Al system, by contrast, forms an L1\textsubscript{2}-ordered intermetallic compound, illustrated in Fig.~\ref{fig:cunial_schematic}, which possesses excellent high-temperature strength. This phase is referred to as the $\gamma'$ phase in the context of Ni-based superalloys~\cite{royer_situ_1998, murakumo_creep_2004}. When Cu and Ni\textsubscript{3}Al are alloyed to form a ternary system, experimental observations confirm that, for Cu-rich compositions (large $x$), L1\textsubscript{2}-ordered Ni\textsubscript{3}Al precipitates out of the disordered ternary solid solution. These precipitates harden the alloy, while leaving behind an almost pure Cu matrix that retains many of the excellent transport properties of elemental Cu~\cite{christofidou_effect_2017, li_cuboidal_2021}. Cu--Ni--Al alloys thus provide an attractive route to combining high strength with high electrical and thermal conductivity, addressing a longstanding challenge in alloy design~\cite{li_promising_2016, zhang_high_2017, fu_strength_2019}. Moreover, from the perspective of development of simulation methods for studying alloy phase transformations and phase equilibria, Cu--Ni--Al serves as an important test case because a single model must capture both ordering and separation tendencies to reproduce the observed experimental behaviour.

Computational modelling approaches for studying solid-solid phase transformations in alloys---providing atomic-scale insight into such phenomena---can be grouped into three broad categories. The first is semi-empirical thermodynamic modelling using methods such as \textsc{CALPHAD}~\cite{miettinen_thermodynamic_2005, turchanin_phase_2007, wang_thermodynamic_2019, davey_first-principles-only_2020, dong_novel_2020, dong_calphad-md_2023}, which parametrise an approximate form of the alloy free energy as a function of phase, composition, and temperature from experimental data and/or first-principles density functional theory (DFT) calculations. The second category is that of atomistic simulation: using appropriate sampling algorithms to determine alloy phase equilibria and infer phase transitions, where the internal energy of the alloy is typically described using either DFT calculations~\cite{eisenbach_first-principles_2019}, some form of interatomic potential (either semi-empirical~\cite{mishin_atomistic_2004} or machine-learned~\cite{rosenbrock_machine-learned_2021}), or a lattice-based model such as a cluster expansion~\cite{woodward_first-principles_2014, shao_accurate_2023, ekborg-tanner_construction_2024}. The third, which we adopt in this work, is based on effective medium theories such as the coherent potential approximation (CPA)~\cite{soven_coherent-potential_1967, gyorffy_coherent-potential_1972, stocks_complete_1978}, which aim to describe the average electronic structure and consequent internal energy of the disordered alloy~\cite{turchi_first-principles_1988, althoff_electronic_1996, ruban_atomic_2004, singh_atomic_2015}. Within such CPA-based methodologies, chemical ordering tendencies and, thus, phase transformations are typically described via some form of concentration wave analysis~\cite{khachaturyan_ordering_1978, gyorffy_concentration_1983}, where it is considered how the internal energy of an alloy responds to spatially inhomogeneous atomic-scale chemical fluctuations.

In this study, to give atomic-scale insight into the aforementioned experimentally observed phenomena, we investigate the phase stability of the technologically relevant Cu--Ni--Al ternary alloy in detail using a combination of \textit{ab initio} electronic structure calculations, a concentration wave analysis, and atomistic Monte Carlo simulations. Our computational workflow successfully reproduces a variety of experimentally observed behaviours at modest computational cost. Specifically, we are able to construct a pseudobinary phase diagram for the $\textrm{Cu}_x(\textrm{Ni}_{3/4}\textrm{Al}_{1/4})_{1-x}$ system, identifying distinct regions where (i) the system forms a fully disordered solid solution, (ii) Cu is soluble in L1\textsubscript{2}-ordered Ni\textsubscript{3}Al, and (iii) L1\textsubscript{2} Ni\textsubscript{3}Al and Cu separate away from one another. Moreover, as the workflow is based on first-principles DFT calculations, we are also able to shed light on the electronic mechanisms driving this behaviour, connecting details of the alloy's electronic structure with a propensity for precipitation, thus providing physical insight into the origin of the observed phase separation into an ordered intermetallic compound and an elemental metal. The present study, therefore, uses the Cu--Ni--Al system to demonstrate a workflow---free from empirically determined parameters---for studying the phenomenon of coherent precipitation in multicomponent solid solutions, with particular relevance to related phenomena such as precipitation strengthening.

The remainder of this work is structured as follows. First, in Sec.~\ref{sec:methods}, we discuss the computational methods used in this work to model the electronic structure and phase behaviour of the considered alloys. Subsequently, in Sec.~\ref{sec:results_discussion}, we present our results, demonstrating that our simulation workflow successfully captures the experimentally observed behaviour of these materials. Additionally, we shed light on the origins of the observed precipitation by considering the electronic structure of selected disordered and ordered phases across a variety of alloy compositions. Finally, in Sec.~\ref{sec:conclusions}, we summarise our findings, venture an outlook on their implications, and suggest potential future avenues of study.

\section{Methods}
\label{sec:methods}

\begin{figure*}[t]
    \centering
    \includegraphics[width=\linewidth]{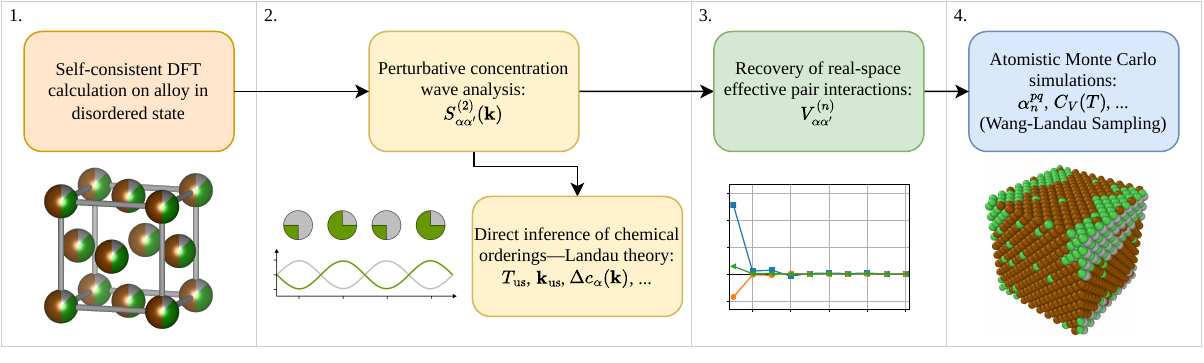}
    \caption{An overview of the computational workflow used in this work for studying the phase behaviour of Cu--Ni--Al alloys.}
    \label{fig:workflow_schematic}
\end{figure*}

Our workflow for studying competing chemical ordering and separation tendencies in substitutional alloys (and, thus, the phenomenon of coherent precipitation in the Cu--Ni--Al system) has four key stages:
\begin{enumerate}
    \item A self-consistent electronic structure calculation is performed with the alloy modelled as a disordered solid solution, as a starting point for the subsequent concentration wave analysis.
    \item A concentration wave analysis applied to the disordered solid solution recovers concentration derivatives of the internal energy, describing atom-atom correlations. A Landau-type mean-field theory then enables direct inference of phase transitions.
    \item From the concentration wave analysis, atom-atom effective pair interactions (EPIs) are recovered for use in subsequent atomistic simulations.
    \item Fixed-lattice Monte Carlo simulations are performed to appropriately sample the configuration space of the alloy, describing phase transformations and phase equilibria in detail.
\end{enumerate}
Many aspects of the approach have been extensively discussed previously~\cite{khan_statistical_2016, woodgate_compositional_2022, woodgate_short-range_2023, woodgate_modelling_2024}, so we present only a concise overview here. A schematic of the workflow is shown in Fig.~\ref{fig:workflow_schematic}, while discussion of the details of each step is provided below. Full computational details of our calculations are provided in Appendix~\ref{appendix:computational_details}.

\subsection{Electronic structure and internal energy of the disordered solid solution}

A substitutional alloy is a material in which there is a fixed underlying crystalline lattice, but in which the arrangements of atoms on said lattice are determined by the laws of statistical physics, \textit{i.e.}\ by thermodynamic considerations. An arrangement of atoms on this underlying lattice can be specified by the set of so-called site occupation numbers, $\{\xi_{i\alpha}\}$, where $\xi_{i\alpha}=1$ if site $i$ is occupied by an atom of chemical species $\alpha$, and $\xi_{i\alpha}=0$ otherwise. Each lattice site is constrained to be occupied by one (and only one) atom, which is equivalent to the statement that $\sum_\alpha \xi_{i\alpha} = 1$ for all lattice sites $i$. Proceeding, the average value of these site occupancies taken across an ensemble defines the site-wise concentrations,
\begin{equation}
    c_{i\alpha} := \langle \xi_{i\alpha} \rangle,
    \label{eq:site-wise_concentrations}
\end{equation}
where $\langle \cdot \rangle$ indicates an average taken over the relevant thermodynamic ensemble. Note that, by construction, we have that $0\leq c_{i\alpha} \leq 1$ for all lattice sites $i$. In the high-temperature limit of this fixed-lattice model, entropy dominates and these site-wise concentrations become spatially homogeneous, \textit{i.e.}
\begin{equation}
    \lim_{T \to \infty} c_{i\alpha} = c_\alpha,
    \label{eq:total_concentrations}
\end{equation}
where $c_\alpha$ is simply the average concentration of chemical species $\alpha$. When the site-wise concentrations, $c_{i\alpha}$, are spatially homogeneous, the disordered alloy is commonly referred to as a `solid solution'. 

The electronic structure and consequent internal energy~\cite{johnson_density-functional_1986, johnson_total-energy_1990} of such a solid solution can be described \textit{ab initio} via application of the coherent potential approximation (CPA)~\cite{soven_coherent-potential_1967, gyorffy_coherent-potential_1972, stocks_complete_1978} within the Korringa--Kohn--Rostoker (KKR) formulation~\cite{korringa_calculation_1947, kohn_solution_1954, ebert_calculating_2011} of DFT~\cite{hohenberg_inhomogeneous_1964, kohn_self-consistent_1965, jones_density_2015}. The KKR method uses multiple scattering theory to obtain the single-particle Green's function for a given solid-state system, while the CPA seeks to construct a self-consistent effective medium of electronic scatterers whose scattering properties represent those of the disordered alloy on average~\cite{faulkner_multiple_2018}. We note that---by its construction---the KKR-CPA typically works with small real-space unit cells (usually the primitive unit cell of the underlying crystal lattice of the alloy), which results in a low computational cost of calculations compared with supercell-based approaches.

\subsection{Describing the energetic cost of atomic-scale chemical fluctuations: Concentration wave analysis}
\label{sec:concentration_waves_methods}

\begin{figure}[t]
    \centering
    \includegraphics[width=\linewidth]{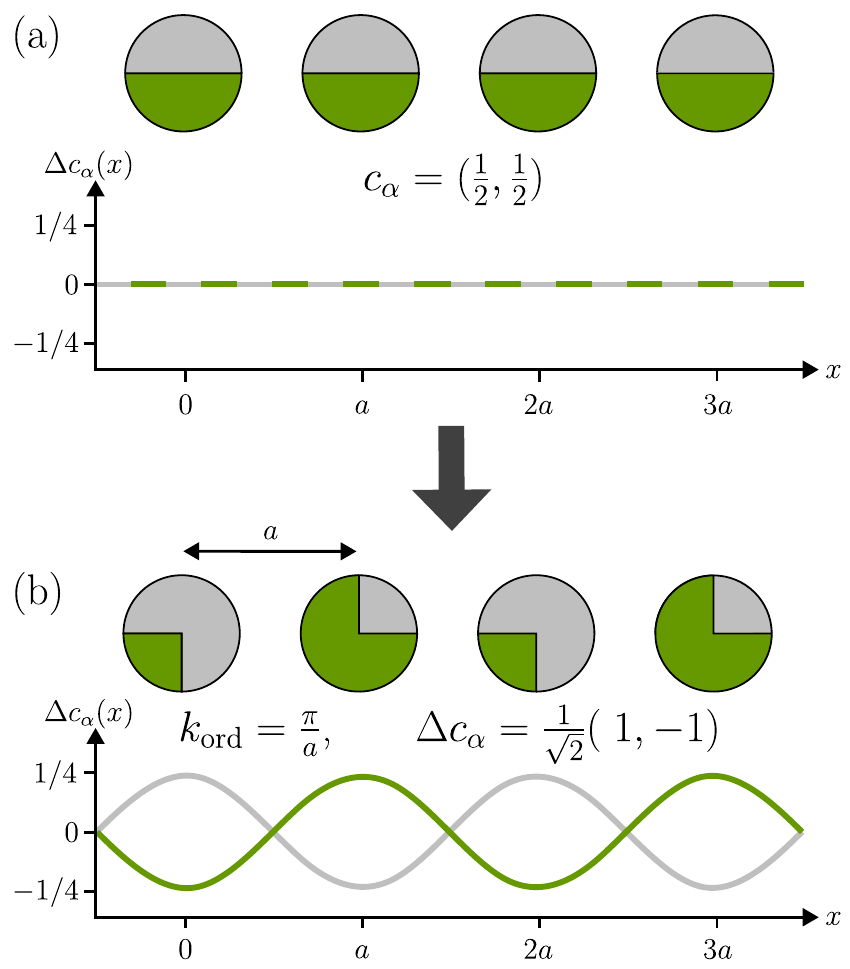}
    \caption{Illustration of a \textit{concentration wave} modulating partial lattice site occupancies in a 1D `alloy'. Such concentration waves can be generalised to three dimensions to describe atomic-scale chemical fluctuations in real alloys. Panel (a) shows the homogeneous alloy with $c_\alpha = (\frac{1}{2}, \frac{1}{2})$, while panel (b) shows those partial occupancies being modulated by a concentration wave with wavevector $k_\textrm{ord} = \pi/a$ and (normalised) chemical polarisation $\Delta c_\alpha = \frac{1}{\sqrt{2}} (1, -1)$. The magnitude of the imposed wave can be scaled arbitrarily so long as $0 \leq c_{i \alpha} \leq 1$ for all lattice sites $i$ and chemical species $\alpha$.}
    \label{fig:concentration_wave_cartoon}
\end{figure}

To study diffusional solid-solid transformations in substitutional alloys---such as the precipitation of L1\textsubscript{2}-ordered Ni\textsubscript{3}Al in Cu--Ni--Al alloys considered in this work---it is necessary to inspect how atomic-scale chemical fluctuations alter the free energy of the alloy. Starting from a disordered solid solution, such chemical fluctuations can be viewed as inhomogeneous perturbations to the homogeneous (disordered) alloy, and we write
\begin{equation}
    c_{i\alpha} = c_{\alpha} + \Delta c_{i\alpha},
    \label{eq:real_space_delta_c}
\end{equation}
where $\Delta c_{i\alpha}$ describes some inhomogeneous atomic-scale chemical fluctuation about the homogeneous reference state. These fluctuations can also be expressed in reciprocal space using the language of \textit{concentration waves}~\cite{khachaturyan_ordering_1978, gyorffy_concentration_1983}, conceptually illustrated in 1D in Fig.~\ref{fig:concentration_wave_cartoon}. In this manner, Eq.~\eqref{eq:real_space_delta_c} is rewritten as
\begin{equation}
    c_{i\alpha} = c_{\alpha} + \sum_{\mathbf{k}} e^{i \mathbf{k} \cdot \mathbf{R}_i} \Delta c_{\alpha}(\mathbf{k}),
\end{equation}
where $\mathbf{R}_i$ is the position of lattice site $i$, $\Delta c_\alpha(\mathbf{k})$ represents some static concentration wave modulating the site-wise concentrations, and the summation typically runs over a handful of high-symmetry $\mathbf{k}$-points.

The change in free energy induced by an arbitrary concentration wave modulating the homogeneous site-wise concentrations is unknown \textit{a priori}, and can be evaluated using two distinct types of approach. The most obvious option is to construct (partially) ordered alloy supercell(s) consistent with a given concentration wave, and evaluate their (free) energy directly. However, for a multicomponent alloy this is laborious due to the large set of possible structures that must be enumerated and, further, incommensurate ordering vectors cannot be treated. An alternative approach, used in this work, is to consider the response of the CPA medium to an infinitesimal concentration fluctuation, and thus to evaluate the concentration derivatives of the alloy internal energy directly. From these concentration derivatives, we construct an expression for the approximate change in free energy (grand potential), $\Delta \Omega$, resulting from deviations from the ideal solid solution, which takes the form
\begin{equation}
    \Delta \Omega = \frac{1}{2} \sum_{\bf k} \sum_{\alpha, \alpha'} \Delta c_\alpha({\bf k}) [\beta^{-1} C^{-1}_{\alpha \alpha'} -S^{(2)}_{\alpha \alpha'}({\bf k})] \Delta c_{\alpha'}({\bf k}),
\label{eq:chemical_stability_reciprocal}
\end{equation}
where $\beta = (k_\mathrm{B} T)^{-1}$ is the usual inverse temperature, $C^{-1}_{\alpha \alpha'} = \delta_{\alpha \alpha'}/c_\alpha$, and $S^{(2)}_{\alpha \alpha'}(\mathbf{k})$ represents the Fourier transform of the second concentration derivative of the alloy internal energy as described within the KKR-CPA, defined formally in Ref.~\citenum{khan_statistical_2016}.

The term appearing in square brackets in Eq.~\eqref{eq:chemical_stability_reciprocal}, $[\beta^{-1} C^{-1}_{\alpha \alpha'} -S^{(2)}_{\alpha \alpha'}({\bf k})]$, we refer to as the \textit{chemical stability matrix}. It can be related to an estimate of the atomic short-range order~\cite{khan_statistical_2016}. In the limit of high temperature (small $\beta$), all its eigenvalues are strictly positive for all wavevectors $\mathbf{k}$. With decreasing temperature, however, it is found that for some temperature $T_\mathrm{us}$ and some (generally high-symmetry) wavevector $\mathbf{k}_\mathrm{us}$ an eigenvalue of this matrix will pass through zero. At this temperature, the free energetic cost of imposing a concentration wave with wavevector $\mathbf{k}_\mathrm{us}$ vanishes, the solid solution is rendered unstable (hence the subscript `us') to such a chemical fluctuation, and a phase transition is inferred. The wavevector associated with this instability describes the structural nature of the transition, while the (normalised) eigenvector, $\Delta c_\alpha$, associated with the eigenvalue passing through zero describes the chemical polarisation of the concentration wave, and thus how the different chemical species partition themselves onto different sublattices. Typically, it is found that the minimum eigenvalue occurs at a high-symmetry point in the Brillouin zone, \textit{i.e.}\ at a special point~\cite{khachaturyan_ordering_1978}. For example, in three dimensions, an instability at the $X$ point, \textit{i.e.}\ with $\mathbf{k}_\mathrm{us} = \frac{2\pi}{a} (0,0,1)$ (and its symmetry-equivalent wavevectors) typically describes chemical ordering into either an L1\textsubscript{0} or L1\textsubscript{2} (sometimes D0\textsubscript{22}) structure~\cite{khachaturyan_ordering_1978, woodgate_structure_2024}, with the final result determined by the system's stoichiometry and free energetic considerations. Phase separation, meanwhile, is described by an instability at the $\Gamma$ point, \textit{i.e.}\ with $\mathbf{k}_\mathrm{us} = \mathbf{0}$~\cite{khachaturyan_ordering_1978, zhang_tailoring_2026}. This analysis represents a Landau-type mean-field theory for inferring phase transitions in an alloy.

It should be emphasised that the alloy $S^{(2)}$ theory used here~\cite{khan_statistical_2016, woodgate_compositional_2022, woodgate_short-range_2023, woodgate_modelling_2024} is a rigorous multicomponent generalisation of an earlier theory restricted to binary alloys~\cite{staunton_compositional_1994, johnson_first-principles_1994, clark_van_1995}. It fully includes the rearrangement of electronic charge due to an inhomogeneous chemical fluctuation, which is crucial for accurately modelling alloys in which there are appreciable charge transfers between constituent elements~\cite{woodgate_compositional_2022, woodgate_short-range_2023}. In previous studies on Al-containing multicomponent alloys~\cite{woodgate_structure_2024, woodgate_emergent_2025, woodgate_electronic_nodate}, we have found that the alloy $S^{(2)}$ theory successfully captures the strong atomic ordering tendencies that emerge when Al is alloyed with $3d$ transition metals, and we therefore believe that this methodology is well-suited to the study of the Cu--Ni--Al alloys considered in the present work.

\subsection{Recovery of a real-space pair interaction model}

In addition to the Landau-type mean-field theory outlined above, used to infer phase transitions directly, it is also possible to map the obtained concentration derivatives of the KKR-CPA internal energy to a real-space atomistic model. The quantity $S^{(2)}_{i\alpha;j\alpha'}$, which is defined as the lattice Fourier transform of $S^{(2)}_{\alpha\alpha'}(\mathbf{k})$, can be related to a set of atom-atom EPIs describing the internal energy of the alloy~\cite{khan_statistical_2016}. This pairwise model, which we refer to as the Bragg--Williams model~\cite{bragg_effect_1934, bragg_effect_1935}, has a generalised Lenz--Ising Hamiltonian, which takes the form
\begin{equation}
    H(\{\xi_{i\alpha}\}) = \frac{1}{2}\sum_{i \alpha; j\alpha'} V_{i\alpha; j\alpha'} \xi_{i \alpha} \xi_{j \alpha'},
    \label{eq:b-w1}
\end{equation}
where $V_{i\alpha; j\alpha'} = -S^{(2)}_{i\alpha;j\alpha'}$ is the EPI between an atom of chemical species $\alpha$ at lattice site $i$ and chemical species $\alpha'$ at lattice site $j$. These EPIs are assumed to be spatially homogeneous (translationally invariant), isotropic (rotationally invariant), and of finite range, which simplifies evaluation of Eq.~\eqref{eq:b-w1} in practical simulations. We then write $V^{(n)}_{\alpha\alpha'}$ to denote the EPI between an atom of chemical species $\alpha$ and an atom of chemical species $\alpha'$ at $n$th-nearest neighbour distance, \textit{i.e.}\ on coordination shell $n$. The gauge degree of freedom for the $\{V_{i\alpha; j\alpha'}\}$ is fixed according to the protocol outlined in Ref.~\citenum{khan_statistical_2016}.

\subsection{Atomistic Monte Carlo simulations}

Using the atom-atom EPIs and the Hamiltonian of Eq.~\eqref{eq:b-w1}, we can subsequently employ lattice-based Monte Carlo simulations for in-depth study of an alloy's phase behaviour. In this work, two algorithms are used: the Metropolis algorithm~\cite{metropolis_equation_1953, landau_guide_2014} and the Wang--Landau sampling algorithm~\cite{wang_efficient_2001, wang_determining_2001, landau_guide_2014}. Here, both of these algorithms make use of Kawasaki-type trial Monte Carlo moves~\cite{kawasaki_diffusion_1966}, \textit{i.e.}\ utilising \textit{swaps} of pairs of atoms in the simulation cell to conserve the overall concentration of each chemical species during the simulation.

The Metropolis algorithm~\cite{metropolis_equation_1953, landau_guide_2014} functions by accepting Monte Carlo moves from an initial state, labelled $n$, to a proposed state, labelled $m$, with probability
\begin{equation}
    P_{n\rightarrow m} = 
    \begin{cases}
        \; \text{exp}\left(-\Delta E/k_\mathrm{B}T\right), &\quad \Delta E > 0\\
        \; 1, &\quad \Delta E \leq 0,
    \end{cases}
\label{eq:metropolis_transition_probability}
\end{equation}
where $\Delta E = E_m - E_n$ is the energy difference between the initial and proposed states, $k_\mathrm{B}$ is the usual Boltzmann constant, and $T$ denotes the simulation temperature. In practice, for the alloys examined in this study, the Metropolis algorithm is used only to obtain equilibrated atomic configurations at a range of temperatures for visualisation purposes. Initially, we prepare a randomly generated supercell containing the specified concentrations of each element and then proceed with the Monte Carlo simulation at a specified temperature. Once a defined number of Monte Carlo steps has been completed, an equilibrated (or near-equilibrated) atomic configuration is obtained and subsequently used for visualisation and potential further study.

In contrast to the Metropolis algorithm, the Wang--Landau (WL) sampling algorithm~\cite{wang_efficient_2001, wang_determining_2001} is a flat-histogram method that can be used to obtain the (configurational) density of states for a given model system, provided it has an associated Hamiltonian that defines the energy as a function of the model's configuration. This (configurational) density of states can then be used to calculate thermodynamic quantities at arbitrary temperature. The WL algorithm works by approximating the partition function, $Z$, as a sum over discretised energy macrostates, \textit{i.e.}
\begin{equation}
    Z \approx \sum_j g(E_j)\, \exp(-E_j / k_\mathrm{B} T),
    \label{eq:macrostate_sum}
\end{equation}
where $g(E)$ is the (discretised, configurational) density of states, and $E_j$ is the energy of the centre of the energy macrostate $j$. An initial guess of the density of states is made (\textit{e.g.}\ $g(E) \equiv 1$) and subsequently refined and converged over the course of the simulation. Further details and discussion of this procedure can be found in Refs.~\citenum{woodgate_emergent_2025}, \citenum{naguszewski_brawl_2025}, and \citenum{naguszewski_optimal_2026}. Once the density of states is known, the corresponding energy probability distribution at temperature $T$ is then defined via
\begin{equation}
    P(E_j, T) = \frac{g(E_j)\, e^{-E_j/k_\mathrm{B} T}}{Z}.
    \label{eq:prob_dist}
\end{equation}
This probability distribution can be used to evaluate a range of thermodynamic quantities. Among these is the configurational specific heat, $C_V$, obtained via~\cite{landau_guide_2014}
\begin{equation}
    C_V(T) = \frac{\langle E^2 \rangle - \langle E \rangle^2}{k_\mathrm{B} T^2},
\end{equation}
where $E$ denotes the system energy, $T$ the temperature, and the ensemble averages, $\langle \cdot \rangle$, are computed using the probability distribution of Eq.~\eqref{eq:prob_dist}.

In addition to the temperature-dependent specific heat, which facilitates identification of the temperatures at which phase transitions occur, we also calculate chemical order parameters to describe the evolution of atomic environments in the simulation as a function of temperature. For this purpose, we employ the Warren--Cowley atomic short-range order (ASRO) parameters~\cite{cowley_short-_1960} adapted to the multicomponent setting, defined as
\begin{equation}
    \alpha^{pq}_n = 1 - \frac{P^{pq}_n}{c_q},
\end{equation}
where $n$ labels the coordination shell under consideration, $P^{pq}_n$ denotes the probability that a site adjacent to an atom of type $p$ on shell $n$ is occupied by species $q$, and $c_q$ is the concentration of species $q$. Values of $\alpha^{pq}_n > 0$ indicate that $p$-$q$ pairings occur less frequently than in a random alloy, whereas $\alpha^{pq}_n < 0$ reflects an enhanced likelihood of such pairings; the case $\alpha^{pq}_n = 0$ corresponds to a random distribution of species $q$ on shell $n$ around species $p$. These short-range order parameters are computed across the energy range sampled in the WL simulation and then recast into their temperature-dependent form using the probability distribution defined in Eq.~\eqref{eq:prob_dist}~\cite{woodgate_emergent_2025}.

\section{Results and Discussion}
\label{sec:results_discussion}

\subsection{DFT-calculated lattice parameters and structural properties}
\label{sec:structural_results}

\begin{figure}[b]
    \centering
    \includegraphics[width=\linewidth]{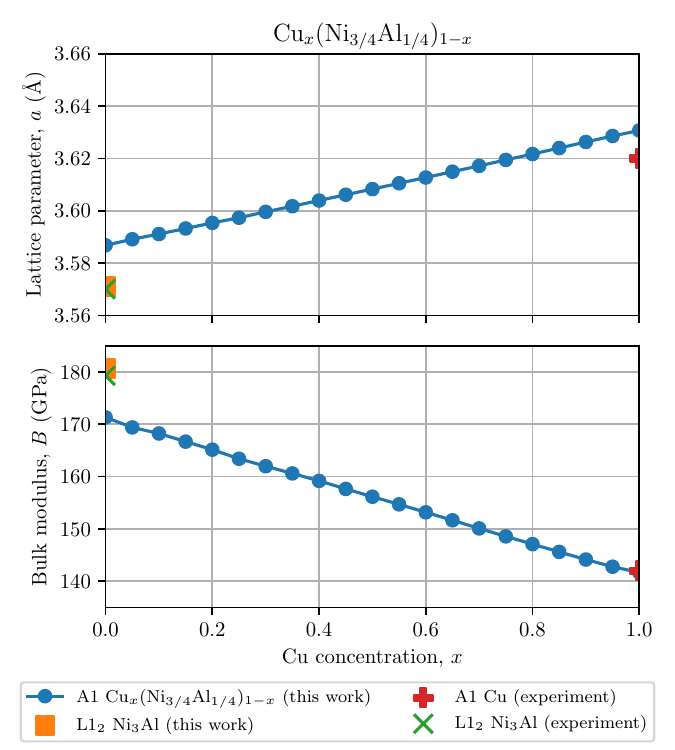}
    \caption{DFT-calculated optimised lattice parameters (top) and bulk moduli (bottom) for an fcc $\textrm{Cu}_x (\textrm{Ni}_{3/4} \textrm{Al}_{1/4})_{1-x}$ solid solution as a function of Cu concentration, $x$. The DFT-optimised lattice parameter increases linearly with $x$, while the bulk modulus decreases. Overall there is good agreement with the highlighted experimental determinations for L1\textsubscript{2}-ordered Ni\textsubscript{3}Al and A1 Cu. It should be noted that the ordered L1\textsubscript{2} phase of Ni\textsubscript{3}Al has a reduced lattice parameter and increased bulk modulus compared with the disordered fcc (A1) phase.}
    \label{fig:structural_properties}
\end{figure}

As outlined above, our workflow begins with self-consistent electronic structure calculations within the KKR formulation~\cite{korringa_calculation_1947, kohn_solution_1954, ebert_calculating_2011} of DFT~\cite{hohenberg_inhomogeneous_1964, kohn_self-consistent_1965, jones_density_2015}, using the CPA~\cite{soven_coherent-potential_1967, gyorffy_coherent-potential_1972, stocks_complete_1978} to average over compositional disorder whenever alloying is considered. Rather than rely on lattice parameters based on experimental determinations, in this work we obtain optimised lattice parameters for the considered systems \textit{ab initio}. In the top panel of Fig.~\ref{fig:structural_properties}, we plot the DFT-optimised lattice parameter calculated for the studied fcc $\textrm{Cu}_x(\textrm{Ni}_{3/4}\textrm{Al}_{1/4})_{1-x}$ solid solutions as a function of $x$. We also include, for comparison, the optimised lattice parameter for L1\textsubscript{2}-ordered Ni\textsubscript{3}Al obtained with the same DFT settings. For elemental Cu, our optimised lattice parameter is $a=3.631$~\AA, in good agreement with the experimental determination of $a=3.615$~\AA~\cite{straumanis_lattice_1969}. Similarly, for L1\textsubscript{2}-ordered Ni\textsubscript{3}Al, our optimised lattice parameter of $a=3.571$~\AA\ compares well with the experimental determination of $a\approx3.572$~\AA~\cite{mishima_lattice_1985}. The agreement is reassuring, but we note here that the PBE exchange-correlation functional~\cite{perdew_generalized_1996} can marginally under-bind metals such as Al, Ni and Cu, and their associated alloys~\cite{grabowski_ab_2007, haas_calculation_2009, csonka_assessing_2009}. It can be seen that there is a modest, linear increase in the DFT-optimised lattice parameter for the system with increasing $x$, in accordance with bulk Cu having a marginally larger lattice parameter than the disordered Ni\textsubscript{3}Al composition endpoint. Additionally, that L1\textsubscript{2} Ni\textsubscript{3}Al has a smaller lattice constant than fcc Cu aligns with the observation that L1\textsubscript{2} Ni\textsubscript{3}Al precipitates can strengthen the alloy, because the mismatch of lattice constants in an otherwise coherent lattice will naturally introduce elastic strain fields in the crystal.

The bottom panel of Fig.~\ref{fig:structural_properties} then shows the DFT-calculated bulk modulus, $B$, as a function of $x$ for the examined Cu--Ni--Al solid solutions. Considering the end points first, we note that the calculated bulk modulus of elemental Cu is $B=141.7$~GPa, in fair agreement with the experimental determination of $B=137.6\pm0.2$~GPa~\cite{ledbetter_elastic_1974}. For the L1\textsubscript{2}-ordered phase of Ni\textsubscript{3}Al, we calculate a bulk modulus of $B=180.9$~GPa, again in fair agreement with the experimental value of $B\approx174$~GPa~\cite{kayser_elastic_1981, prikhodko_temperature_1999}. Proceeding, inspecting the trends observed across the $\textrm{Cu}_x(\textrm{Ni}_{3/4}\textrm{Al}_{1/4})_{1-x}$ pseudobinary system, it can be seen that the addition of Cu to the solid solution reduces the calculated bulk modulus of the alloy, which decreases linearly with increasing $x$. The calculated bulk modulus of L1\textsubscript{2} Ni\textsubscript{3}Al is substantially higher than that of elemental Cu, confirming that the ordered Ni\textsubscript{3}Al precipitates constitute a markedly stiffer phase than the Cu-rich matrix. Taking $B$ as a proxy for the overall elastic stiffness of a material, this also supports the claim that these precipitates can induce modulus mismatch strengthening in the alloy.

\subsection{Concentration wave analysis}
\label{sec:results_concentration_wave_analysis}

\begin{figure*}[t]
    \centering
    \includegraphics[width=\linewidth]{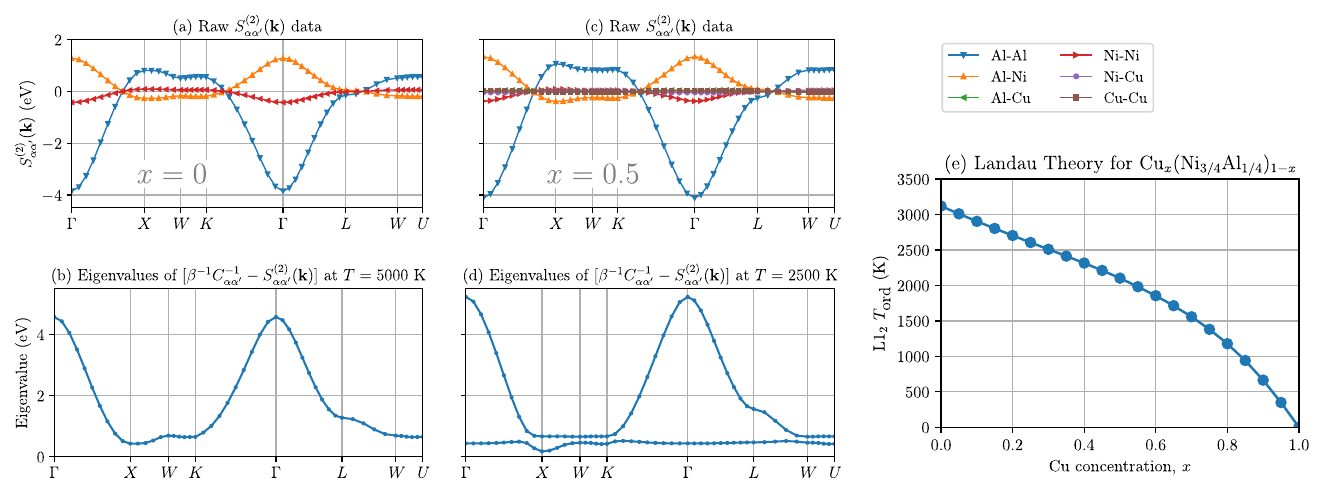}
    \caption{Data associated with the \textit{ab initio} concentration wave analysis performed on the considered $\textrm{Cu}_x(\textrm{Ni}_{3/4}\textrm{Al}_{1/4})_{1-x}$ pseudobinary system. Panels (a) and (b) show, respectively, raw $S^{(2)}_{\alpha \alpha'}(\mathbf{k})$ data and eigenvalues of the chemical stability matrix for the case $x=0$, \textit{i.e.}\ the binary Ni\textsubscript{0.75}Al\textsubscript{0.25}. Panels (c) and (d) then show the same data for the case $x=0.5$. Finally, panel (e) shows L1\textsubscript{2} ordering temperatures as a function of $x$ predicted by the mean-field Landau-type theory outlined in Sec.~\ref{sec:methods}.}
    \label{fig:s2_theory_results}
\end{figure*}

To assess the (free) energetic cost of atomic-scale chemical fluctuations in the solid solution, we then use the self-consistent one-electron potentials and associated electron density as a starting point for a concentration wave analysis, as outlined in Sec.~\ref{sec:methods}. For the compositions $x=0$ and $x=0.5$, in Fig.~\ref{fig:s2_theory_results} we present plots of the quantity $S^{(2)}_{\alpha \alpha'}(\mathbf{k})$ along high-symmetry lines of the irreducible Brillouin zone of the fcc lattice, together with eigenvalues of the chemical stability matrix along those same high-symmetry lines. That there are one (two) eigenvalues of the chemical stability matrix at each $\mathbf{k}$-point for a two-component (three-component) alloy follows from the restriction that any given chemical fluctuation must preserve the overall concentration of each chemical species~\cite{khan_statistical_2016}.

We consider first the raw $S^{(2)}_{\alpha \alpha'}(\mathbf{k})$ data, which can be thought of as the lattice Fourier transform of the real-space atom-atom EPIs. For the case $x=0$, \textit{i.e.}\ the composition Ni$_{0.75}$Al$_{0.25}$ for which the relevant data are shown in panels (a) and (b) of Fig.~\ref{fig:s2_theory_results}, we see that the largest in magnitude, most strongly varying values of $S^{(2)}_{\alpha \alpha'}(\mathbf{k})$ are those corresponding to Al-Al and Ni-Al interactions, while those associated with Ni-Ni pairs are the smallest in magnitude and more weakly varying. The minimum eigenvalue of the chemical stability matrix lies at the $X$ point, \textit{i.e.}\ $\mathbf{k}=\frac{2\pi}{a}(0,0,1)$, and has associated (normalised) eigenvector $ \Delta c_\alpha = \frac{1}{\sqrt{2}}(1, -1)$. (Note that, for a two-component alloy, the eigenvector has two components, while for a three-component alloy, the eigenvector has three components.) This combination is typically indicative of an L1\textsubscript{2}-like ordering for a system with the stoichiometry $A_{0.75}B_{0.25}$~\cite{khachaturyan_ordering_1978, woodgate_modelling_2024}, though we re-emphasise that distinguishing between L1\textsubscript{2}-like and D0\textsubscript{22}-like order is a subtle point~\cite{johnson_first-principles_1994, woodgate_structure_2024}; this is settled by the Monte Carlo simulations presented below.

For the case $x=0.5$, \textit{i.e.}\ the composition $\textrm{Cu}_{0.5}\textrm{Ni}_{0.375} \textrm{Al}_{0.125}$ for which the relevant data are shown in panels (c) and (d) of Fig.~\ref{fig:s2_theory_results}, the $S^{(2)}_{\alpha \alpha'}(\mathbf{k})$ data reveal a similar picture for Al-Al, Ni-Al, and Ni-Ni pairs. The data associated with Cu, \textit{i.e.}\ for Al-Cu, Ni-Cu, and Cu-Cu pairs, are much smaller in magnitude than for other pairs, which we associate with comparatively weak interactions between Cu and the other two elements (Ni, Al) in this system. As for $x=0$, the minimum eigenvalue of the chemical stability matrix lies at the $X$ point, again indicative of L1\textsubscript{2}-like ordering, though this time the associated (normalised) eigenvector is $ \Delta c_\alpha = (0.7134,\,-0.7006,\, -0.0128)$. (Throughout, when chemical polarisations are given, components of the vector associated with different elements are consistently ordered from smallest to largest atomic number: $\alpha=1$ corresponds to Al, $\alpha=2$ corresponds to Ni, and $\alpha=3$ corresponds to Cu~\footnote{Note also that, wherever eigenvectors of the chemical stability matrix are quoted in the text, we provide their numerical value at the temperature at which their associated eigenvalue passes through zero. With increasing temperature there may be some modest rotation of these vectors in concentration space.}.) Combined with the fact that the minimum eigenvalue is at the $X$ point, this eigenvector is best interpreted in terms of strong ordering between Ni and Al atoms, with little or no redistribution of Cu atoms. We note also that there is no clear competing minimum eigenvalue at $\Gamma$ indicating phase separation. This suggests that the separation of Cu from L1\textsubscript{2} Ni\textsubscript{3}Al is a secondary instability. Again, though, this aspect is captured in the Monte Carlo simulations which follow. Overall, then, the picture that emerges is one where the dominant atom-atom interactions in the system involve Ni and Al atoms, while Cu atoms appear largely inert at this level and play a diluting role. The electronic origins of this behaviour will be discussed further in Sec.~\ref{sec:electronic_origins}.

\begin{figure*}[t]
    \centering
    \includegraphics[width=\linewidth]{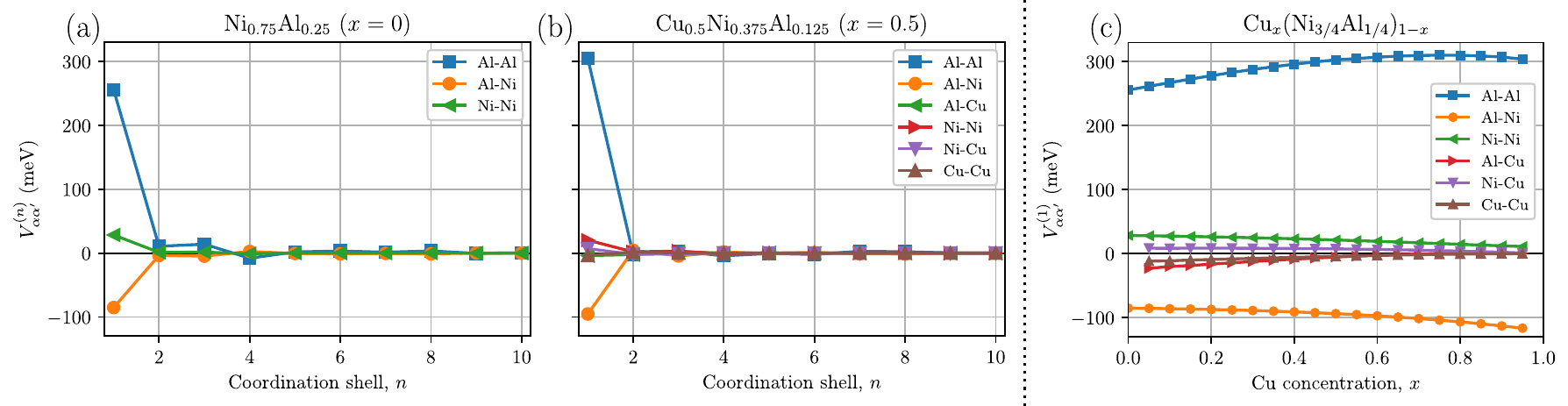}
    \caption{Selected calculated atom-atom effective pair interactions for the $\textrm{Cu}_x (\textrm{Ni}_{3/4} \textrm{Al}_{1/4})_{1-x}$ pseudobinary system. Panels (a) and (b) show $V^{(n)}_{\alpha \alpha'}$ as a function of coordination shell, $n$, for  $x=0$ and $x=0.5$, respectively. Panel (c) then shows $V^{(1)}_{\alpha \alpha'}(x)$, \textit{i.e.}\ just the nearest-neighbour effective pair interactions as a function of Cu concentration, $x$. Interactions are dominated by nearest-neighbour pairs, and they vary smoothly as a function of Cu concentration, $x$. Of course, in panel (c), interactions involving Cu are not defined for $x=0$, while no interactions are formally defined in the case of pure Cu, $x=1$.}
    \label{fig:EPIs}
\end{figure*}

Continuing, as outlined in Sec.~\ref{sec:methods}, a mean-field estimate of an alloy's chemical ordering temperature can be obtained by seeking the temperature at which the lowest-lying eigenvalue of the chemical stability matrix passes through zero, indicating the instability of the solid solution to an applied chemical perturbation represented by the eigenvector associated with that eigenvalue. The temperature at which this instability occurs for the considered alloys is plotted as a function of $x$ in panel (e) of Fig.~\ref{fig:s2_theory_results}. Across all considered compositions, we find that the minimum eigenvalue consistently occurs at the $X$ point, $\mathbf{k}=\frac{2\pi}{a}(0,0,1)$, indicative of an L1\textsubscript{2}-like chemical ordering, while the associated eigenvector (defining the chemical polarisation of the associated concentration wave) remains broadly consistent and typically indicates Al and Ni moving to separate sublattices, with Cu remaining disordered. For example, for $x=0.25$ the eigenvector is $ \Delta c_\alpha =(0.7197, \, -0.6938, \, -0.0259)$, while for $x=0.75$ the eigenvector is $\Delta c_\alpha = (0.6611,\,  -0.7455,\, 0.0844)$. We re-emphasise, though, that this Landau-type theory only infers the \textit{dominant} chemical ordering and that, in a system such as Cu--Ni--Al where both chemical ordering and phase separation are known to occur, it will be necessary to go further than this mean-field analysis to accurately describe the phase diagram, as discussed below.

In terms of comparison with previous studies, the case $x=0$, \textit{i.e.}\ the Ni$_{0.75}$Al$_{0.25}$ binary, presents the most obvious test case. Here our mean-field Landau-type theory provides an estimate of the (virtual) ordering temperature~\footnote{In this context, a `virtual' ordering refers to a hypothetical chemical ordering that takes place above the melting temperature of the material.} of $T_\mathrm{ord} \approx 3100$~K, much higher than the (virtual) experimental value~\cite{cahn_order-disorder_1987} of $T_\mathrm{ord} = 1720 \pm 10$~K. This overestimate is, however, expected. First, we emphasise that the chemically ordered L1\textsubscript{2} phase is stable up to the melting point experimentally~\cite{bremer_order-disorder_1988, okamoto_al-ni_2004}, and we believe that our lattice-based model predicting an ordering temperature higher than the alloy's melting temperature is consistent with these experimental data. Moreover, as mentioned above, we have not included the effects of vibrational entropy in this analysis~\cite{van_de_walle_effect_2002}. Inclusion of such effects would be anticipated to reduce the predicted chemical ordering temperature, because this is typically found to be the case when an ordered phase is elastically stiffer than its precursor disordered phase, as has previously been found to be the case in Ni\textsubscript{3}Al~\cite{ravelo_free_1998, michelon_computational_2010}. In particular, we note that Ref.~\citenum{michelon_computational_2010} reports a 30\% reduction in estimated ordering temperature for this system following the inclusion of such effects. Finally, we stress that a mean-field analysis, such as the Landau-type theory utilised here, is likely to overestimate true ordering temperatures.

\subsection{Atom-atom effective pair interactions}
\label{sec:results_epis}

From the reciprocal-space $S^{(2)}_{\alpha\alpha'}(\mathbf{k})$ data---representing the second variation of the energy of the KKR-CPA reference medium with respect to inhomogeneous, atomic-scale chemical fluctuations---we subsequently recover real-space atom-atom EPIs, which enable us to assess the strength and range of interactions between all pairs of chemical species. As illustrative examples, in panels (a) and (b) of Fig.~\ref{fig:EPIs}, we plot the recovered atom-atom EPIs, $V^{(n)}_{\alpha \alpha'}$, as a function of coordination shell number, $n$, for the cases $x=0$ and $x=0.5$. It can be seen that the EPIs are relatively short-ranged, being dominated by contributions at nearest-neighbour distance, $n=1$. (That being said, non-negligible interactions persist beyond nearest-neighbour distance, which is most clearly seen for $x=0$, panel (a) of Fig.~\ref{fig:EPIs}.) The strongest interactions are between Al-Al pairs, which are heavily disfavoured on the first coordination shell (positive value of $V^{(1)}_{\alpha \alpha'}$). The second strongest interactions are then between Ni and Al pairs, which are favoured on the first coordination shell (negative value of $V^{(1)}_{\alpha \alpha'}$). We note that these nearest-neighbour interactions are consistent with the pair preferences observed in the L1\textsubscript{2} crystal structure. By comparison with the Al-Al and Ni-Al interactions, Ni-Ni pairs and---where relevant---Al-Cu, Ni-Cu, and Cu-Cu pairs are all found to interact comparatively weakly. Given the absence of long-ranged interactions, we choose to truncate our fits, discussed in Appendix~\ref{appendix:computational_details}, to the first six coordination shells of the fcc lattice for use in the fixed-lattice Monte Carlo simulations that follow. We note that this truncation corresponds, for example, to a real-space `cut-off' of $r_\mathrm{cut}=6.2$~\AA\ for $x=0$, comparable with typical real-space cut-offs used in other simulation approaches, \textit{e.g.}\ for machine-learned interatomic potentials~\cite{zhang_roadmap_2025}.

\begin{figure*}[t]
    \centering
    \includegraphics[width=\linewidth]{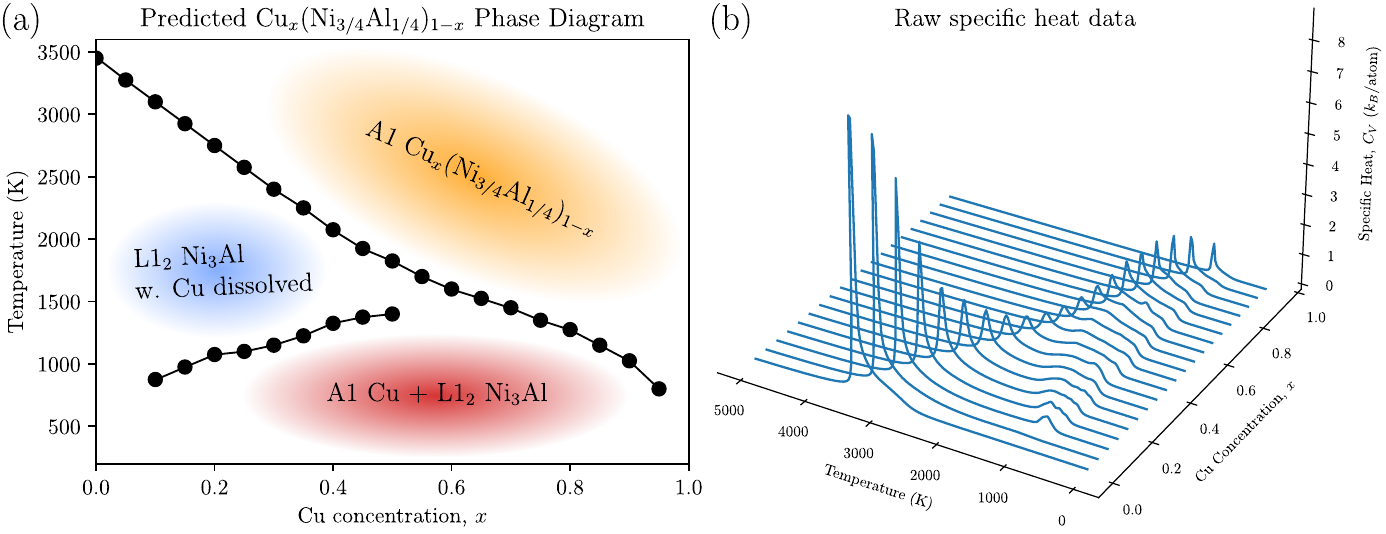}
    \caption{Overview of the inferred pseudobinary phase diagram (a), with phase boundaries recovered from the raw specific heat data obtained from fixed-lattice Monte Carlo simulations using the Wang--Landau sampling algorithm (b). Boundaries on the phase diagram are positioned according to the location of local peaks in the specific heat data. We emphasise that this is a fixed-lattice configurational diagram neglecting melting and vibrational free energy contributions.}
    \label{fig:wang-landau_data}
\end{figure*}

The smooth and systematic variation of the recovered atom-atom EPIs as a function of varying Cu concentration, $x$, is evidenced in panel (c) of Fig.~\ref{fig:EPIs}, where we plot $V^{(1)}_{\alpha \alpha'}(x)$, \textit{i.e.}\ the nearest-neighbour interactions for all elemental pairings as a function of $x$. In alignment with the above observations for $x=0$ and $x=0.5$, across the entire considered composition space, it is consistently found that Al-Al nearest-neighbour interactions are strongest, Ni-Al next-strongest, and other pair interactions comparatively weak. That the EPIs vary smoothly as a function of composition suggests that it may be possible in future to sample a limited number of points in composition space and subsequently fit a smooth function and interpolate between them to recover EPIs for arbitrary compositions. This smoothness, therefore, is an encouraging observation in the context of potential future explorations of high-dimensional compositional spaces.

\subsection{Monte Carlo simulations}
\label{results:monte_carlo}

\begin{figure*}[t]
    \centering
    \includegraphics[width=\linewidth]{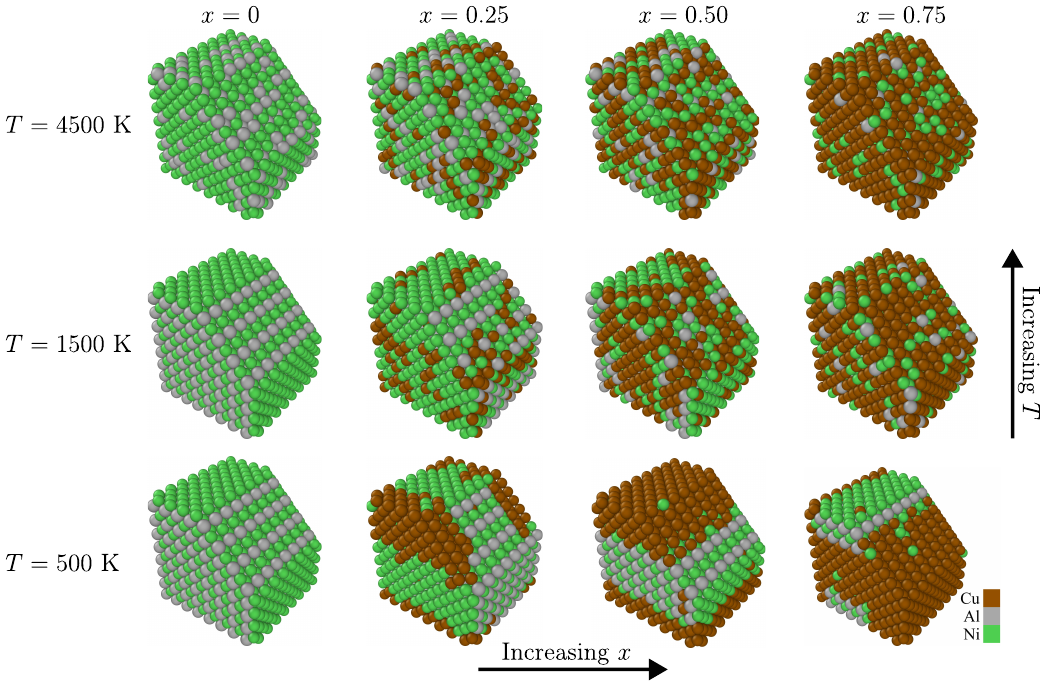}
    \caption{Indicative equilibrium atomic arrangements for the $\textrm{Cu}_x (\textrm{Ni}_{3/4} \textrm{Al}_{1/4})_{1-x}$ pseudobinary system as a function of Cu concentration, $x$, and temperature, $T$ as generated using fixed-lattice Monte Carlo simulations. Though $T=4500$~K is above the melting temperature of the alloy, we visualise these configurations here to confirm that our simulations recover a fully disordered alloy in the limit of high temperature. Structural images generated using \textsc{Ovito}~\cite{stukowski_visualization_2010}.}
    \label{fig:monte_carlo_configurations}
\end{figure*}

Using the recovered real-space atom-atom EPIs, a subset of which were visualised in Fig.~\ref{fig:EPIs}, we subsequently perform fixed-lattice Monte Carlo simulations using both the Wang--Landau sampling and Metropolis algorithms to analyse the phase stability of the system in detail. Panel (a) of Fig.~\ref{fig:wang-landau_data} shows a pseudobinary phase diagram for the $\textrm{Cu}_x (\textrm{Ni}_{3/4} \textrm{Al}_{1/4})_{1-x}$ system in the $(x,T)$ plane. This phase diagram is inferred from the raw specific heat data obtained from the WL simulations shown in panel (b) of Fig.~\ref{fig:wang-landau_data}, and from the temperature-dependent Warren--Cowley ASRO parameters plotted in the Supplemental Material~\cite{supplemental}. In support of these data, and for ease of interpretation of our results, in Fig.~\ref{fig:monte_carlo_configurations} we also show indicative equilibrium atomic configurations at selected temperatures and values of $x$ obtained using the Metropolis algorithm. These Monte Carlo simulation data present a more nuanced picture than the results obtained using the mean-field theory in Sec.~\ref{sec:results_concentration_wave_analysis} above. There are three distinct identifiable regimes, and we discuss each in turn. We re-emphasise here that our modelling neglects the effects of differences in vibrational entropy between phases~\cite{van_de_walle_effect_2002, ravelo_free_1998, michelon_computational_2010} and so ordering temperatures will likely be overestimated. Nonetheless, we believe the relevant experimental trends are qualitatively well captured and that the modelling give insights into the physical mechanisms at play.

First, at low Cu concentrations ($x \lesssim 0.1$), there is a single identifiable (virtual) phase transition, taking place above the alloy's melting temperature. In the case $x=0$ this is simply the transition from the disordered $\gamma$ phase (A1) to the ordered $\gamma'$ phase (L1\textsubscript{2}). In the case $x=0.05$, the lack of a second identifiable peak in the specific heat data suggests a single phase transition from the $\gamma$ to $\gamma'$ phase of Ni\textsubscript{3}Al, with Cu soluble in both phases. We note that the question of L1\textsubscript{2}-like versus D0\textsubscript{22}-like ordering is fully resolved by these simulations; from our visualised configurations and from the Warren--Cowley ASRO parameters depicted in the Supplemental Material~\cite{supplemental}, it is clear that our modelling predicts L1\textsubscript{2}-like order.

Second, at intermediate Cu concentrations ($0.1 \lesssim x \lesssim 0.5$), there are two identifiable phase transitions, with two associated peaks in the specific heat data. When the specific heat data are combined with the calculated Warren--Cowley ASRO parameters visualised in the Supplemental Material~\cite{supplemental}, we are able to infer that the sharp, high-temperature peak is associated with the formation of L1\textsubscript{2}-ordered Ni\textsubscript{3}Al while Cu remains soluble. The second, smaller, broader peak in the specific heat data we then associate with phase separation, where elemental Cu clusters away from L1\textsubscript{2}-ordered Ni\textsubscript{3}Al at lower temperatures. This assessment is also supported by the visualised indicative equilibrium atomic configurations for the case $x=0.25$ shown in Fig.~\ref{fig:monte_carlo_configurations}. Here it can be seen that at $T=1500$~K Cu mixes with L1\textsubscript{2}-ordered Ni\textsubscript{3}Al, while at $T=500$~K the two phases have separated.

Third, at higher Cu concentrations ($0.5 \lesssim x \lesssim 1$), there is only one identifiable phase transition with decreasing temperature. Here, supported by the indicative atomic configurations shown in Fig.~\ref{fig:monte_carlo_configurations}, we infer that L1\textsubscript{2}-ordered Ni\textsubscript{3}Al (the $\gamma'$ phase) precipitates directly out of the solid solution. It should be emphasised here that this direct precipitation is advantageous from a materials properties perspective, because the case in which Cu is soluble in L1\textsubscript{2} Ni\textsubscript{3}Al (occurring at small values of $x$ in our modelling) results in a material with residual disorder and consequently poor transport properties. By contrast, when Ni\textsubscript{3}Al precipitates directly out of the solid solution leaving near-pure Cu behind (as is the case for higher Cu concentrations in our modelling), this near-pure Cu matrix will retain many of the good transport properties associated with elemental Cu~\cite{li_cuboidal_2021, li_strong_2021, li_resistivity-temperature_2025}.

Qualitatively, these results are broadly in agreement with the experimentally reported pseudobinary phase diagram of Semboshi \textit{et al.}~\cite{semboshi_phase_2022}. These authors find that Cu is soluble in L1\textsubscript{2} Ni\textsubscript{3}Al (in their considered temperature range) up to a maximum of around 10\% Cu, \textit{i.e.}\ $x=0.1$, in alignment with our own findings. It should be stressed, though, that their boundary between the two-phase region (miscibility gap) and the region in which Cu is soluble in L1\textsubscript{2} Ni\textsubscript{3}Al is shifted relative to ours in the $(x,T)$ plane, appearing at higher temperatures and not extending to as large values of $x$ at a given temperature as ours does. At larger values of $x$, these authors also report a direct transition from the fcc solid solution to the two-phase region, albeit at lower temperatures than ours. For example, at $x=0.8$, these authors report a transition at a temperature of approximately $750$~$^\circ$C (approximately 1020~K) while we find a transition at around $1250$~K, \textit{i.e.}\ approximately 980~$^\circ$C.

\subsection{Electronic origins of ordering and separation}
\label{sec:electronic_origins}

\begin{figure}[b]
    \centering  \includegraphics[width=\linewidth]{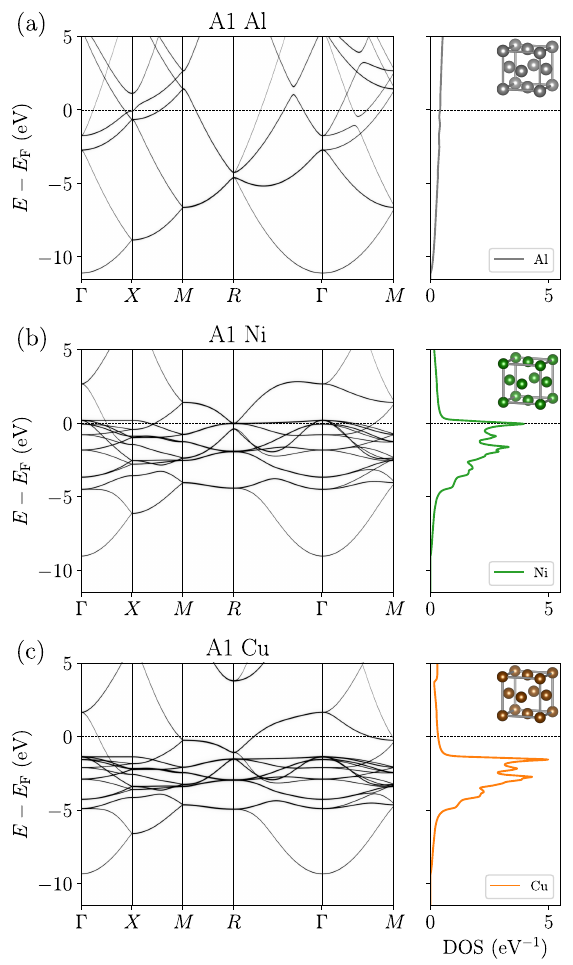}
    \caption{Bloch spectral function (band structure) and electronic density of states for elemental fcc (A1) Al (a), Ni (b), and Cu (c). The parabolic, nearly-free-electron-like bands of Al should be contrasted with the narrower $3d$ bands of the transition metals Ni and Cu. Structural images (insets) generated using \textsc{VESTA}~\cite{momma_vesta_2011}.}\label{fig:elemental_electronic_structure}
\end{figure}

\begin{figure*}[t]
    \centering  \includegraphics[width=\textwidth]{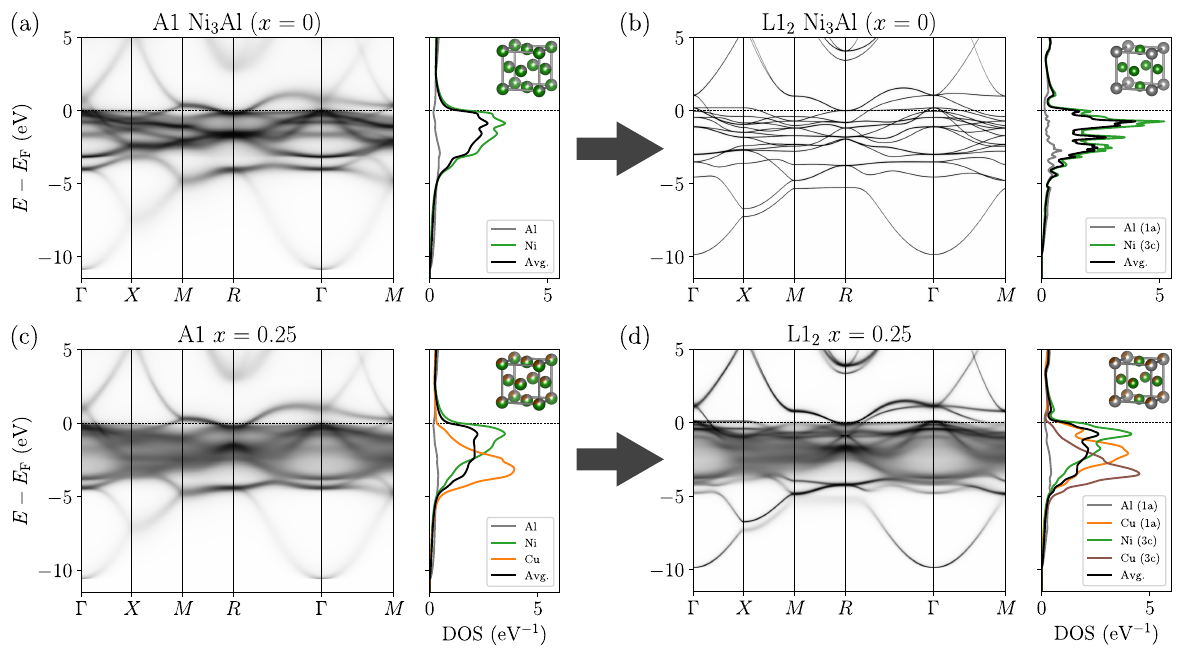}
    \caption{Details of the evolving electronic structure of the $\textrm{Cu}_x (\textrm{Ni}_{3/4} \textrm{Al}_{1/4})_{1-x}$ pseudobinary system as a result of L1\textsubscript{2} chemical ordering. Panels (a) and (b) show the BSF (band structure) and electronic DOS for Ni\textsubscript{3}Al in disordered fcc (A1) and L1\textsubscript{2}-ordered states. Panel (c) then shows the case $x=0.25$ modelled in a disordered fcc (A1) state. Finally, panel (d) shows the case $x=0.25$ in which L1\textsubscript{2} order has become established between Ni and Al, but where Cu remains homogeneously spread (dissolved) across the system. Structural images (insets) generated using \textsc{VESTA}~\cite{momma_vesta_2011}.}\label{fig:ordering_electronic_structure}
\end{figure*}

We now examine details of the electronic structures of the constituent elements, disordered alloys, and (partially) ordered intermetallic compounds, to elucidate the underlying electronic origins of the strong chemical ordering tendencies in the $\textrm{Cu}_x (\textrm{Ni}_{3/4} \textrm{Al}_{1/4})_{1-x}$ pseudobinary system. We carry out calculations of the Bloch spectral function (BSF), $A_\mathrm{B} ({\bf k}, E)$ for this purpose.  Formally the BSF is given in terms of the single-particle Green's function which sits at the heart of the KKR method~\cite{ebert_calculating_2011}. For an ordered system at $T=0$~K the BSF effectively reduces to the usual band structure 
\begin{equation}
A_\mathrm{B} (\mathbf{k}, E) = \sum_n \delta (E - E_n({\bf k})),
\end{equation}
\textit{i.e.} a set of Dirac delta functions which map out the electronic bands, $E = E_n(\mathbf{k})$. For disordered systems, however, ensemble averages of the Green's function are taken so that the Bloch spectral function typically resembles a set of Lorentzian type-functions, where the loci of the peak positions describe bands, while the $E$ and ${\bf k}$ dependent broadening captures the effects of disorder.  In the figures following presenting results for the $\textrm{Cu}_x (\textrm{Ni}_{3/4} \textrm{Al}_{1/4})_{1-x}$ pseudobinary alloys, the BSF data therefore resemble smeared band structures.  In these terms we describe the electronic structure of both the disordered (A1) solid solution phase, which has an underlying fcc lattice, as well as the (partially) L1\textsubscript{2}-ordered Ni\textsubscript{3}Al-based systems. We present all calculations using the simple-cubic representation of the fcc lattice, with an associated 4-atom basis. In this representation, all four sub-lattices are equivalent for the fcc solid solution (A1 phase), whereas the systems with L1\textsubscript{2}-type order have two distinct sublattices, with Wyckoff labels 1a (corners) and 3c (face centres).

In Fig.~\ref{fig:elemental_electronic_structure}, for reference, we show the electronic structures of the elemental fcc (A1) metals, Al, Ni, and Cu in panels (a), (b), and (c), respectively. (Here we note that elemental Cu represents the $x=1$ limit of the $\textrm{Cu}_x (\textrm{Ni}_{3/4} \textrm{Al}_{1/4})_{1-x}$ system.) Fig.~\ref{fig:ordering_electronic_structure} then shows how the electronic structures of the disordered alloys evolve during (partial) chemical ordering. Panels (a) and (b) show results for solid solution A1 and L1\textsubscript{2}-ordered Ni\textsubscript{3}Al, respectively, \textit{i.e.}\ the case $x=0$. Panels (c) and (d) show results for the case $x=0.25$. Panel (c) shows this system modelled as disordered fcc (A1), while panel (d) shows this system once some ordering has taken place, so that Al has moved to the 1a site, Ni to the 3c sites, and Cu is uniformly dissolved in L1\textsubscript{2}-ordered Ni\textsubscript{3}Al. Throughout, the BSF is evaluated along high-symmetry lines of the simple-cubic Brillouin zone. In the case of an fcc system, this means that electronic bands are shown folded back into the simple-cubic Brillouin zone. Note that, for these systems with multiple elements in the composition, we plot both the element-resolved DOS and the `average' DOS per atom, denoted `Avg.', which is simply the total DOS for the 4-atom simulation cell divided by four.

We start our analysis with reference to Fig.~\ref{fig:elemental_electronic_structure} which shows results for the elemental metals.  
Elemental Al, as a `simple' metal with no associated $d$ electrons, has nearly free electron-like, broadly parabolic $sp$-bands and a DOS which is approximately proportional to the square root of energy measured from the band minimum. These features are consistent with Al's good electrical transport properties. Elemental Ni has a more complex band structure, with a collection of flat, weakly dispersive $3d$-bands located around and below the Fermi level, $E_\mathrm{F}$ as well as more dispersive $sp$-bands.  The large value of the DOS at $E_\mathrm{F}$ for Ni when modelled as non-magnetic, as in this work, ensures that elemental Ni fulfils the Stoner criterion for ferromagnetism~\cite{stoner_collective_1938}. Since our calculations for the pseudobinary systems are chosen to be non-magnetic because we are focussed explicitly on modelling these alloys' phase stability at temperatures above $T_\textrm{C}$ we use non-magnetic elemental Ni's electronic structure as a useful reference. (As discussed in Appendix~\ref{appendix:magnetism}, self-consistent DLM calculations for both A1 and L1\textsubscript{2}-ordered Ni\textsubscript{3}Al found that Ni atoms did not support local magnetic moments in the paramagnetic regime.) Finally, elemental Cu has a similar $3d$ complex to Ni, but one shifted to lower energies on account of the $3d$ states being completely filled in this metal. Around $E_\mathrm{F}$, the bands are more dispersive and $sp$-like in character. As it is the availability and nature of states around $E_\mathrm{F}$ that dominate electrical transport properties, these dispersive bands are consistent with the excellent electrical transport properties of elemental Cu.

Continuing, we examine the electronic structure of the two selected disordered (A1) solid solutions, panels (a) and (c) of Fig.~\ref{fig:ordering_electronic_structure}, which correspond to $x=0$ and $x=0.25$ $\textrm{Cu}_x (\textrm{Ni}_{3/4} \textrm{Al}_{1/4})_{1-x}$ systems respectively. For the case $x=0$, the disordered (A1) alloy, \textit{i.e.} Ni$_{0.75}$Al$_{0.25}$, we note three key features. First, the band structure is heavily smeared across the entire Brillouin zone on account of the compositional disorder of the solid solution and the differing electronic scattering properties of Al and Ni. Second, though the BSF is smeared and there are fewer sharp features in the DOS compared to elemental Ni, the narrow $3d$ bands are still clearly identifiable. Third, a low-lying, nearly-free-electron-like band is pulled down in energy relative to elemental Ni, signalling the onset of Al($p$)--Ni($d$) bonding states. This is the disordered alloy precursor to the strong $p$--$d$ hybridisation we identify below for ordered Ni\textsubscript{3}Al. 

When the case $x=0.25$ is considered, many of the observations for the case $x=0$ carry over. The principal difference, however, is that the Ni and Cu $3d$ complexes lie at markedly different energies relative to $E_\mathrm{F}$. The consequence is a lack of $d$-$d$ bonding states formed by the two late transition metals which hampers the process of mixing the two metals together. This is similar to the species-resolved DOS of Cu--Ni alloys~\cite{seib_photoemission_1970, seib_photoemission_1970-1, stocks_complete_1978} in which pronounced short-range clustering of the Ni atoms is found~\cite{mozer_neutron_1968, vrijen_clustering_1978}. The strong disorder from this sizeable energy separation between the Ni and Cu $3d$ states leads to a combined $3d$ spectral weight which is smeared more heavily than in A1 Ni$_{0.75}$Al$_{0.25}$.

We now examine how the electronic structures of these alloys change upon chemical ordering, considering first panel (b) of Fig.~\ref{fig:ordering_electronic_structure}, which shows results for the perfectly ordered L1\textsubscript{2} Ni\textsubscript{3}Al intermetallic. Its band structure is sharply defined and the two non-equivalent lattice 3c and 1a sites in the unit cell lead to some bands that (although smeared out) were present in the disordered A1 phase having their degeneracy lifted. 
A key feature here is how the element-resolved DOS for Al differs in this ordered intermetallic phase compared to that for elemental Al. Whereas in elemental Al the DOS was smooth and broadly $\propto \sqrt{E}$ for some shifted energy scale, here there are now many sharp features in the element-resolved DOS for Al. These peaks and troughs coincide with the peaks and troughs for the element-resolved DOS for Ni, and confirm strong $p$--$d$ hybridisation in this material. These $p$--$d$ Al-Ni bonding states shift to lower energies in the ordered L1\textsubscript{2} phase compared to the A1 phase. These electronic features explain why our DFT-calculated energy (enthalpy) calculations find that L1\textsubscript{2}-ordered Ni\textsubscript{3}Al's energy is 315~meV/atom lower than that of the same composition modelled as disordered fcc (A1). We emphasise here that this is a zero-temperature enthalpy difference evaluated within the KKR-CPA, which neglects the local lattice relaxations in the disordered phase that would act to reduce this difference. Nonetheless, the size of this enthalpy difference, together with the strong $p$--$d$ hybridisation that produces it, is fully consistent with the experimentally established stability of the ordered phase up to the melting point~\cite{cahn_order-disorder_1987, bremer_order-disorder_1988, okamoto_al-ni_2004}.

Panel (d) of Fig.~\ref{fig:ordering_electronic_structure} shows the result of a calculation performed for the case $x=0.25$ in which it is assumed that L1\textsubscript{2} order has become fully established between Ni and Al, but where Cu remains evenly dispersed throughout the system. This is consistent both with the initial transition predicted by our concentration wave analysis, and with the intermediate-temperature region of the phase diagram predicted at small $x$ values by the Monte Carlo simulations. Here the element-resolved DOS is provided separately for Cu atoms sitting on the Al (1a) sites and Ni (3c) sites; the two environments are different and so too is the site-resolved DOS. When the BSF data is considered, it can be seen that, at some portions of the Brillouin zone, there is significantly reduced smearing on account of the order established between Ni and Al atoms, while at others the remaining disorder coming from the Ni and Cu atoms still smears out fine details of the band structure, particularly at energies where the $3d$ bands make sizeable contributions to the DOS. Additionally, while the DOS contribution from Cu on the 3c site is largely unaltered by the ordering, the $3d$-derived DOS contribution from Cu on the 1a site is pushed up towards $E_\mathrm{F}$ coinciding with a Ni 3c DOS contribution. This Ni-Cu bonding feature increases the contribution to the total energy made by Cu atoms as compared to a situation in which they are able to separate out of the L1\textsubscript{2}-ordered phase. Hence, while there is a reduction in energy-per-atom going from the fully disordered phase, described in panel (c), to the partially ordered phase described in panel (d) which is caused primarily by the Al-Ni $p$--$d$ bonding and calculated as  183~meV/atom, the energy reduction going from the disordered phase, described in panel (c), to two separate phases of elemental Cu and L1\textsubscript{2}-ordered Ni\textsubscript{3}Al is larger still at 259~meV/atom. That the reduction in energy from a fully disordered phase to an L1\textsubscript{2}-ordered Ni\textsubscript{3}Al phase with Cu dissolved (183~meV/atom) is larger than the further reduction in energy caused by Cu separating out ($259-183 = 76$~meV/atom) we interpret as corroborating our earlier finding that the L1\textsubscript{2} ordering is the initial, high-temperature transition in the small-$x$ regime, with eventual separation into near-pure Cu and L1\textsubscript{2}-ordered Ni\textsubscript{3}Al being a secondary transition occurring at lower temperatures.

\section{Summary and Conclusions}
\label{sec:conclusions}

In summary, an analysis of the phase stability and electronic structure of ternary Cu--Ni--Al alloys, specifically the $\textrm{Cu}_x (\textrm{Ni}_{3/4} \textrm{Al}_{1/4})_{1-x}$ pseudobinary system, has been presented. The modelling approach employed a combination of first-principles electronic structure calculations, a perturbative concentration wave analysis (from which atom-atom effective pair interactions were recovered), and fixed-lattice atomistic Monte Carlo simulations using both the Metropolis and Wang--Landau sampling algorithms.

It was shown that---in the considered fcc solid solutions---Al and Ni atoms interact strongly with one another, while Cu interacts more weakly and plays a diluting role. Three distinct regimes of phase behaviour were revealed: At small values of $x$, Cu was found to be soluble in L1\textsubscript{2}-ordered Ni\textsubscript{3}Al, with the single identifiable phase transition being the high-temperature (virtual) ordering between Ni and Al. At intermediate values of $x$, two separate phase transitions were identified: the first being an ordering into an L1\textsubscript{2} structure with Cu dissolved, and the second being phase separation between L1\textsubscript{2}-ordered Ni\textsubscript{3}Al and an fcc Cu-rich phase. Finally, at larger values of $x$, the single identifiable phase transition was the precipitation of L1\textsubscript{2}-ordered Ni\textsubscript{3}Al directly from the ternary solid solution, \textit{i.e.}\ the ordering and phase separation occur concurrently. These results were found to compare qualitatively favourably with the available experimental data, albeit with the fixed-lattice modelling understood to result in overestimation of ordering temperatures. Further, analysis of the electronic structure of several of the considered solid solutions gave insight into the underlying electronic origins of these ordering and separation tendencies. Specifically, strong $p$-$d$ hybridisation tendencies between Al and Ni were confirmed to be the driver for chemical ordering between these species. Cu, with its lower-lying $3d$ bands compared with Ni, is then expelled from the ordered phase of Ni\textsubscript{3}Al at lower temperatures and higher Cu concentrations.

Overall, this work serves as a demonstration of a computationally efficient method, free of empirically determined parameters, for studying composition-dependent phase behaviour in multicomponent alloys in which there is a coherent underlying lattice. In particular, these results demonstrate that the methodology is applicable to the study of the phenomenon of coherent precipitation in multicomponent systems, where a unified modelling framework must be capable of describing both chemical ordering and phase separation tendencies simultaneously. The approach is therefore relevant to the study of related phenomena such as precipitation strengthening. Moreover, because our method is based on first-principles electronic structure calculations, it remains possible to interpret ordering and separation tendencies in terms of an alloy's evolving electronic structure. Potential extensions of this work in the context of this class of alloy include developing means by which to rigorously include the effects of vibrational entropy differences between phases throughout the simulation methodology, as well as extending the predicted phase diagram towards the full Cu--Ni--Al ternary. More generally, future studies could seek to use this approach to screen for new alloy compositions with desirable physical properties for particular applications.

\section*{Acknowledgments}
C.D.W. acknowledges support from an UK Engineering and Physical Sciences Research Council (EPSRC) Doctoral Prize Fellowship at the University of Bristol, Grant EP/W524414/1. H.J.N. acknowledges a funded PhD studentship within the EPSRC Centre for Doctoral Training in Modelling of Heterogeneous Systems, Grant EP/S022848/1. J.M. thanks the project QM4ST, funded as project No. CZ.02.01.01/00/22$\_$008/0004572 by Programme Johannes Amos Commenius, call Excellent Research. J.B.S. acknowledges support from EPSRC Grant EP/W021331/1. Compute resources were provided by the Advanced Computing Research Centre at the University of Bristol and by the Scientific Computing Research Technology Platform of the University of Warwick. J.M. also acknowledges VSB – Technical University of Ostrava, IT4Innovations National Supercomputing Center, Czech Republic, for awarding this project access to the LUMI supercomputer, owned by the EuroHPC Joint Undertaking, hosted by CSC (Finland) and the LUMI consortium through the Ministry of Education, Youth and Sports of the Czech Republic through the e-INFRA CZ (grant ID: 90254).

C.D.W. and J.B.S. conceived of the approach, with input from all authors.  C.D.W. performed self-consistent electronic structure calculations, the concentration wave analysis, and fitting of the atom-atom effective pair interactions, with support from J.M. and J.B.S. H.J.N. implemented the Wang--Landau sampling algorithm in the \textsc{Brawl} package and performed Monte Carlo sampling simulations using both the Metropolis algorithm and Wang--Landau sampling under the supervision of D.Q., with support from C.D.W. S.L.D. and M.P.P. worked jointly on electronic structure calculations of the relevant disordered and ordered alloys during their final-year undergraduate project at the University of Warwick. C.D.W. prepared the first draft of the manuscript with support from H.J.N. Subsequently, all authors revised the manuscript and approved its final version.

\appendix

\section{Computational details}
\label{appendix:computational_details}

Full details of our calculations---including all relevant simulation input and output files---can be found in the dataset associated with this work~\cite{dataset}, but for convenience here we summarise the key computational details. Throughout, we sample a total of 21 evenly spaced values of $x$, \textit{i.e.}\ adjusting the Cu concentration by 5\% between each value of $x$, to describe the $\textrm{Cu}_x(\textrm{Ni}_{3/4}\textrm{Al}_{1/4})_{1-x}$ pseudobinary system.

\subsection{Electronic structure calculations}

\subsubsection{DFT parameters}

For an accurate description of the structural properties of these alloys, we perform scalar-relativistic full-potential (\textit{i.e.}\ without shape restriction on the atom-centred potentials) calculations using the all-electron \textsc{SPR-KKR} package~\cite{ebert_calculating_2011, sprkkr_website}, which implements the KKR formulation~\cite{korringa_calculation_1947, kohn_solution_1954, faulkner_multiple_2018} of DFT. Chemical disorder is described via application of the CPA~\cite{soven_coherent-potential_1967, gyorffy_coherent-potential_1972, stocks_complete_1978}. The generalised gradient approximation (GGA) to the DFT exchange-correlation functional is used, specifically the parametrisation of Perdew, Burke, and Ernzerhof (PBE)~\cite{perdew_generalized_1996}. Integrations over valence energies are performed using a 32-point semi-circular contour in the complex plane, and an angular momentum cut-off of $l_\textrm{max}=4$ is employed. For Brillouin zone integrations, we use the parameter \texttt{NKTAB=1000}, which results in a dense $36 \times 36 \times 36$ $\mathbf{k}$-point mesh over the first Brillouin zone of the fcc lattice. These calculations are not spin-polarised; justification of this choice is given below, in Sec.~\ref{appendix:magnetism}. Optimised lattice parameters are obtained by performing self-consistent calculations at a series of lattice parameters to obtain the energy as a function of cell volume for a given composition. Subsequently, these data are fitted using a stabilised jellium equation of state (SJEOS)~\cite{alchagirov_energy_2001} as implemented in the Atomic Simulation Environment (ASE)~\cite{hjorth_larsen_atomic_2017}, from which an optimised lattice parameter and the bulk modulus can be recovered.

Following the optimisation of lattice parameters at a given composition, we freeze the lattice parameter at this value and perform all remaining calculations within the atomic sphere approximation (ASA)~\cite{andersen_linear_1975} using the local density approximation (LDA) to the DFT exchange-correlation functional as parametrised by Vosko, Wilk, and Nusair (VWN)~\cite{vosko_accurate_1980}, with an angular momentum cut-off of $l_\textrm{max}=3$. All other calculation parameters are unchanged from the calculations outlined above. These approximations---LDA, ASA---are necessitated, at present, by the complexity of the concentration wave analysis outlined above, in Sec.~\ref{sec:concentration_waves_methods}. We emphasise that we have previously found that the results of the concentration wave analysis are not strongly affected by small variations of lattice parameter because the analysis depends on relative energies and band-filling trends~\cite{woodgate_competition_2024, woodgate_modelling_2024}.

\subsubsection{Treatment of magnetism}
\label{appendix:magnetism}

In an alloy containing a magnetic element such as Ni, due care must be taken with regard to an appropriate description of an alloy's magnetic state when simulating its phase behaviour~\cite{woodgate_interplay_2023, ghosh_chemical_2024}. Though elemental Ni---the magnetic ground state of which is ferromagnetic---has a comparatively high Curie temperature of $T_\mathrm{C} \sim 630$~K~\cite{kouvel_detailed_1964}, once alloyed with non-magnetic elements---such as Cu or Al---this value falls rapidly. For example, binary Cu$_y$Ni$_{1-y}$ alloys have a Curie temperature that decreases approximately linearly with increasing $y$, and are found to be non-magnetic for $y > 0.524$~\cite{ododo_critical_1977}. Similarly, when Ni is alloyed with Al, L1\textsubscript{2}-ordered Ni\textsubscript{3}Al has a measured Curie temperature (at ambient pressure) of approximately 41~K, though this value depends strongly on the precise stoichiometry~\cite{de_boer_exchange-enhanced_1969, hamid_electronic_2011}. These data establish that the relevant magnetic state to consider when simulating phase equilibria in the $\textrm{Cu}_x(\textrm{Ni}_{3/4}\textrm{Al}_{1/4})_{1-x}$ pseudobinary system is paramagnetic, leaving open whether that paramagnet is best described by persistent, atomically localised moments or by itinerant electrons with no stable local moment. To distinguish between these two pictures, we attempted a self-consistent, spin-polarised disordered-local-moment (DLM) calculation~\cite{staunton_disordered_1984} for the disordered (fcc) and ordered (L1\textsubscript{2}) phases of the composition with the highest Ni concentration ($x=0$), at the obtained DFT-optimised lattice parameters. The local moments collapsed during self-consistency, indicating a paramagnetic phase largely itinerant in character, consistent with the weakly localised moments of elemental Ni~\cite{staunton_using_2014}. Throughout this work, therefore, the studied alloys are treated as non-magnetic.

\subsection{Concentration wave analysis}

An in-house implementation of the $S^{(2)}$ theory for multicomponent alloys is used to perform the concentration wave analysis. This code makes use of an adaptive meshing scheme for Brillouin zone integrations~\cite{bruno_algorithms_1997}. The quantity $S^{(2)}_{\alpha \alpha'}(\mathbf{k})$, which varies smoothly as a function of $\mathbf{k}$, is evaluated for each composition at a total of 56 points of the irreducible Brillouin zone of the fcc lattice, including high-symmetry points. Practically, the real-space effective pair interactions of Eq.~\eqref{eq:b-w1} are recovered via a least-squares fitting procedure to the reciprocal-space data. (Further discussion of this fitting procedure can be found in Sec.~II.B of Ref.~\citenum{woodgate_short-range_2023}.)

\subsection{Monte Carlo simulations}

The Monte Carlo simulations within this study were performed using the \textsc{Brawl} package~\cite{naguszewski_brawl_2025}; the parallelisation strategies employed for the WL sampling algorithm are discussed in Ref.~\citenum{naguszewski_optimal_2026}. Within our WL simulations, the density of states histograms for $\textrm{Cu}_x (\textrm{Ni}_{3/4} \textrm{Al}_{1/4})_{1-x}$ were obtained using a system of $8\times 8 \times 8$ fcc cubic unit cells consisting of a total of 2048 atoms in an initially random configuration with periodic boundary conditions. The applied WL $\log(f)$ tolerance~\cite{wang_efficient_2001, wang_determining_2001} was $1\cdot10^{-10}$ with a desired flatness of 85\%. All simulations utilised 1024 uniform energy bins across the corresponding energy range for the composition.

\end{document}


\title{Formation of L1\textsubscript{2}-ordered $\gamma'$-Ni\textsubscript{3}Al precipitates in ternary Cu--Ni--Al alloys modelled using an \textit{ab initio} concentration wave theory and atomistic simulations\\\vspace{6pt}Supplemental Material}

\author{Christopher D. Woodgate}
\thanks{Corresponding author}
\email{Christopher.Woodgate@bristol.ac.uk}
\affiliation{H.H. Wills Physics Laboratory, University of Bristol, Royal Fort, Bristol, BS8 1TL, United Kingdom}
\affiliation{Department of Physics, University of Warwick, Coventry, CV4 7AL, United Kingdom}
\author{Hubert J. Naguszewski}
\affiliation{Department of Physics, University of Warwick, Coventry, CV4 7AL, United Kingdom}
\author{Samuel L. Deacon}
\thanks{These authors contributed equally to this work.}
\affiliation{Department of Physics, University of Warwick, Coventry, CV4 7AL, United Kingdom}
\author{Mathys P. Potel}
\thanks{These authors contributed equally to this work.}
\affiliation{Department of Physics, University of Warwick, Coventry, CV4 7AL, United Kingdom}
\author{Ján Minár}
\affiliation{New Technologies Research Center, University of West Bohemia, Pilsen, Czech Republic}
\author{David Quigley}
\affiliation{Department of Physics, University of Warwick, Coventry, CV4 7AL, United Kingdom}
\author{Julie B. Staunton}
\email{J.B.Staunton@warwick.ac.uk}
\affiliation{Department of Physics, University of Warwick, Coventry, CV4 7AL, United Kingdom}

\maketitle

This is the Supplemental Material accompanying the main text. Here, we present combined plots of the recovered specific heat, $C_V$, and Warren-Cowley atomic short-range order (ASRO) parameters, $\alpha_{n}^{pq}$, as a function of temperature, $T$, for each alloy composition. These data serve as evidence justifying the pseudobinary phase diagram presented in the main text.

\section{Monte Carlo simulation results}

\begin{figure}[hp]
    \centering
    \includegraphics[width=\linewidth]{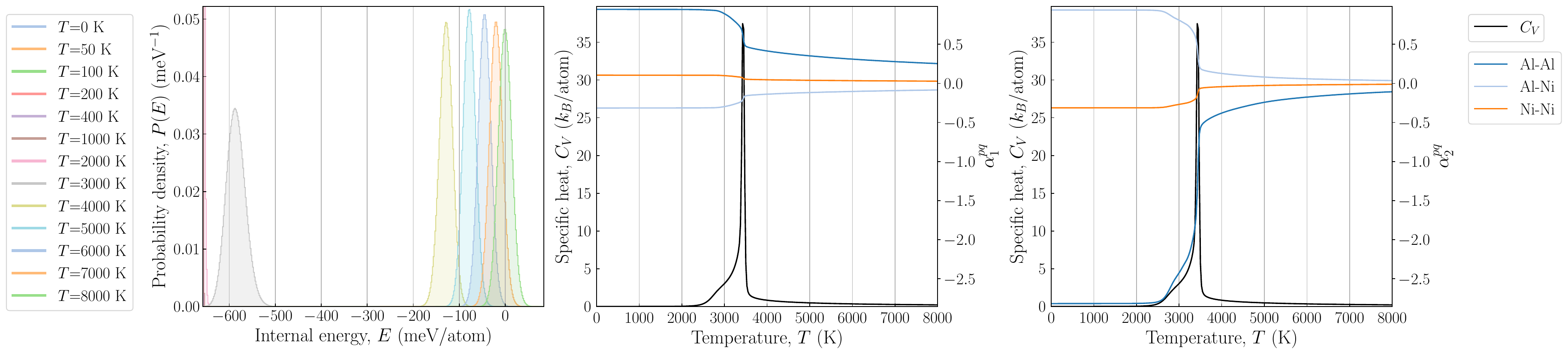}
    \caption{Obtained specific heat ($C_V$) and Warren-Cowley ASRO parameters ($\alpha^{n}_{pq}$) as a function of temperature for Ni$_{0.75}$Al$_{0.25}$, \textit{i.e.} $\textrm{Cu}_x(\textrm{Ni}_{3/4}\textrm{Al}_{1/4})_{1-x}$ for $x=0$}
    \label{fig:x_0.00}
\end{figure}

\begin{figure}[hp]
    \centering
    \includegraphics[width=\linewidth]{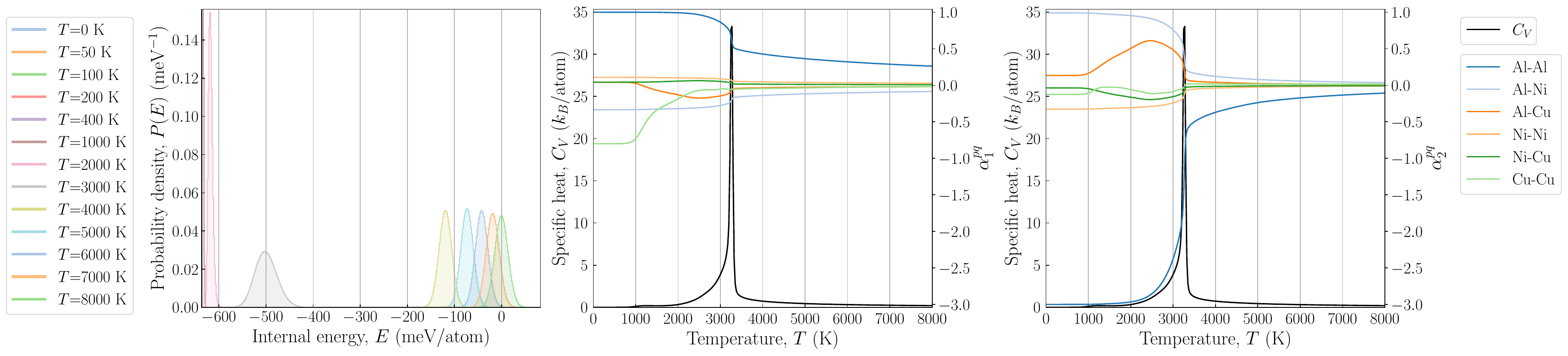}
    \caption{Obtained specific heat ($C_V$) and Warren-Cowley ASRO parameters ($\alpha^{n}_{pq}$) as a function of temperature for Cu$_{0.05}$Ni$_{0.7125}$Al$_{0.2375}$, \textit{i.e.} $\textrm{Cu}_x(\textrm{Ni}_{3/4}\textrm{Al}_{1/4})_{1-x}$ for $x=0.05$}
    \label{fig:x_0.05}
\end{figure}

\begin{figure}[hp]
    \centering
    \includegraphics[width=\linewidth]{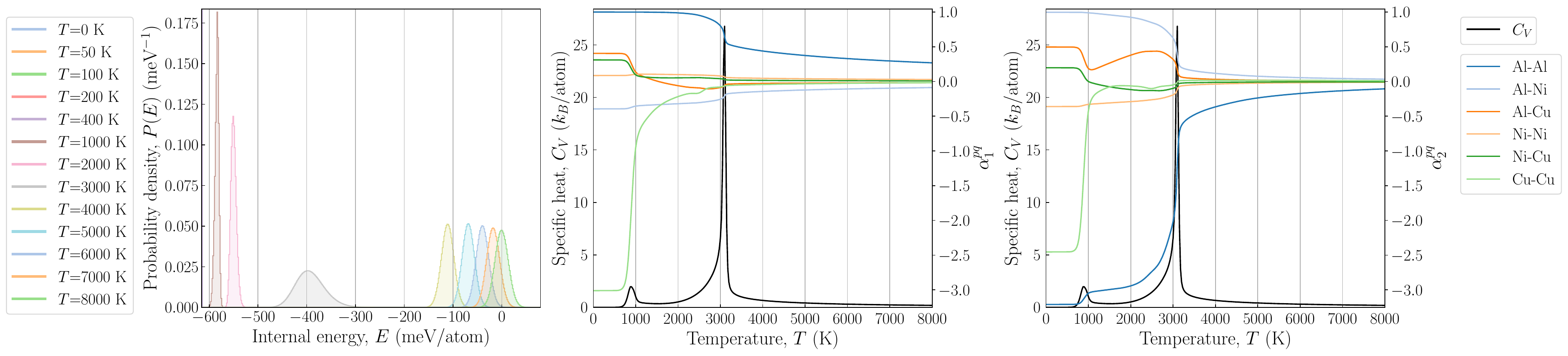}
    \caption{Obtained specific heat ($C_V$) and Warren-Cowley ASRO parameters ($\alpha^{n}_{pq}$) as a function of temperature for Cu$_{0.1}$Ni$_{0.675}$Al$_{0.225}$, \textit{i.e.} $\textrm{Cu}_x(\textrm{Ni}_{3/4}\textrm{Al}_{1/4})_{1-x}$ for $x=0.1$}
    \label{fig:x_0.10}
\end{figure}

\begin{figure}[hp]
    \centering
    \includegraphics[width=\linewidth]{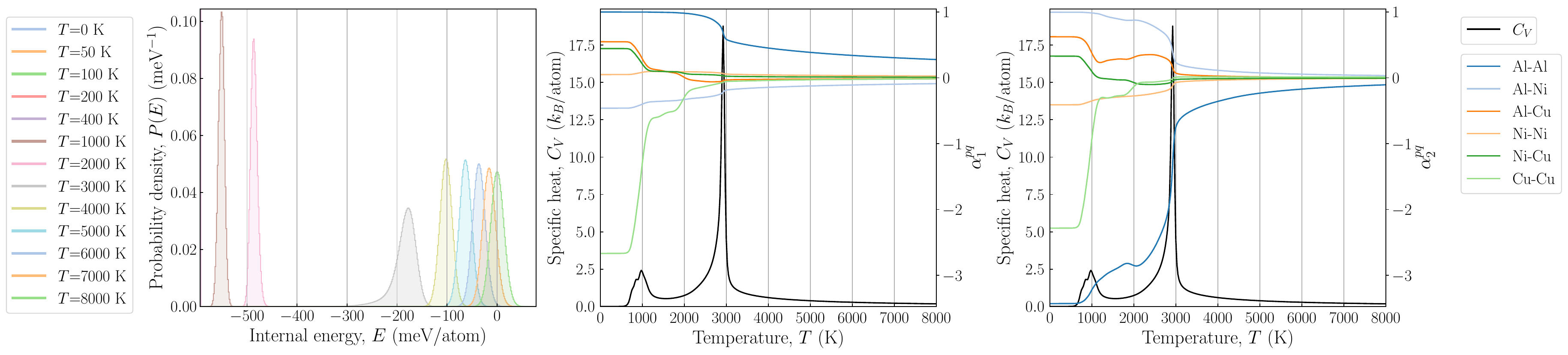}
    \caption{Obtained specific heat ($C_V$) and Warren-Cowley ASRO parameters ($\alpha^{n}_{pq}$) as a function of temperature for Cu$_{0.15}$Ni$_{0.6375}$Al$_{0.2125}$, \textit{i.e.} $\textrm{Cu}_x(\textrm{Ni}_{3/4}\textrm{Al}_{1/4})_{1-x}$ for $x=0.15$}
    \label{fig:x_0.15}
\end{figure}

\begin{figure}[hp]
    \centering
    \includegraphics[width=\linewidth]{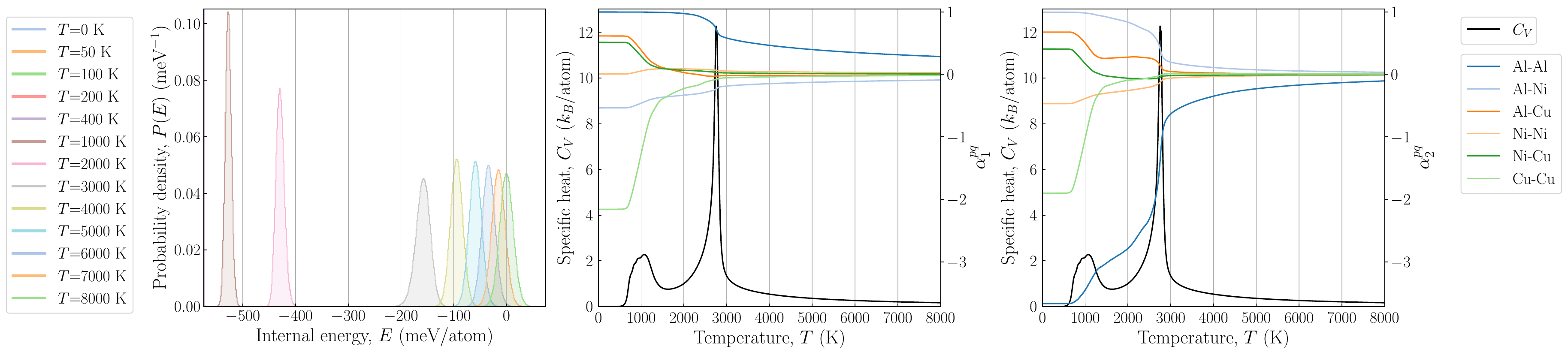}
    \caption{Obtained specific heat ($C_V$) and Warren-Cowley ASRO parameters ($\alpha^{n}_{pq}$) as a function of temperature for Cu$_{0.2}$Ni$_{0.6}$Al$_{0.2}$, \textit{i.e.} $\textrm{Cu}_x(\textrm{Ni}_{3/4}\textrm{Al}_{1/4})_{1-x}$ for $x=0.2$}
    \label{fig:x_0.20}
\end{figure}

\begin{figure}[hp]
    \centering
    \includegraphics[width=\linewidth]{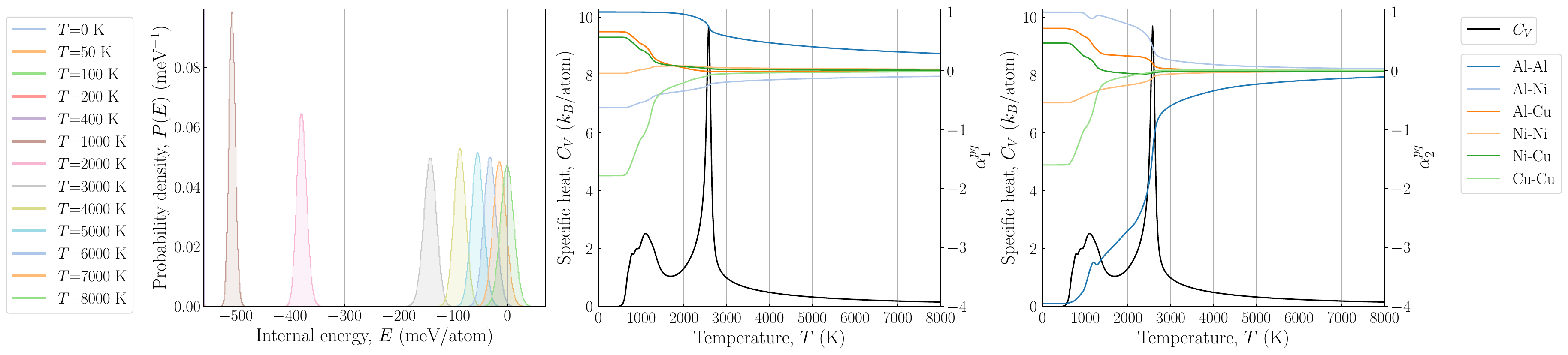}
    \caption{Obtained specific heat ($C_V$) and Warren-Cowley ASRO parameters ($\alpha^{n}_{pq}$) as a function of temperature for Cu$_{0.25}$Ni$_{0.5625}$Al$_{0.1875}$, \textit{i.e.} $\textrm{Cu}_x(\textrm{Ni}_{3/4}\textrm{Al}_{1/4})_{1-x}$ for $x=0.25$}
    \label{fig:x_0.25}
\end{figure}

\begin{figure}[hp]
    \centering
    \includegraphics[width=\linewidth]{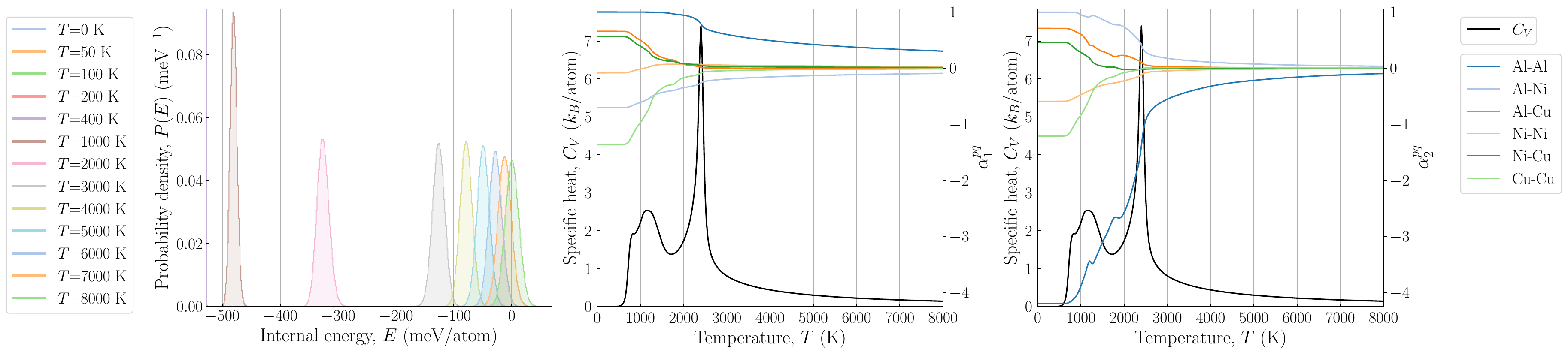}
    \caption{Obtained specific heat ($C_V$) and Warren-Cowley ASRO parameters ($\alpha^{n}_{pq}$) as a function of temperature for Cu$_{0.3}$Ni$_{0.525}$Al$_{0.175}$, \textit{i.e.} $\textrm{Cu}_x(\textrm{Ni}_{3/4}\textrm{Al}_{1/4})_{1-x}$ for $x=0.3$}
    \label{fig:x_0.30}
\end{figure}

\begin{figure}[hp]
    \centering
    \includegraphics[width=\linewidth]{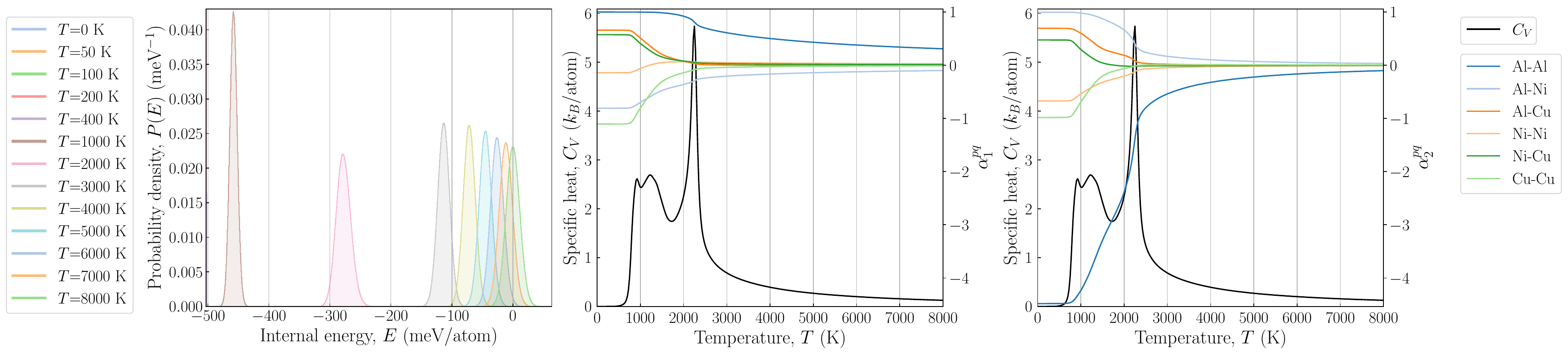}
    \caption{Obtained specific heat ($C_V$) and Warren-Cowley ASRO parameters ($\alpha^{n}_{pq}$) as a function of temperature for Cu$_{0.35}$Ni$_{0.4875}$Al$_{0.1625}$, \textit{i.e.} $\textrm{Cu}_x(\textrm{Ni}_{3/4}\textrm{Al}_{1/4})_{1-x}$ for $x=0.35$}
    \label{fig:x_0.35}
\end{figure}

\begin{figure}[hp]
    \centering
    \includegraphics[width=\linewidth]{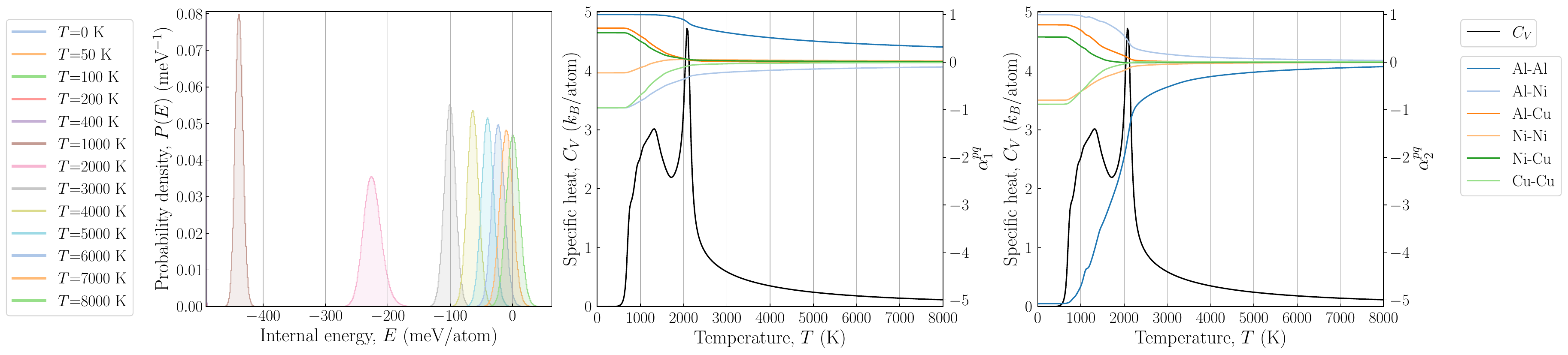}
    \caption{Obtained specific heat ($C_V$) and Warren-Cowley ASRO parameters ($\alpha^{n}_{pq}$) as a function of temperature for Cu$_{0.4}$Ni$_{0.45}$Al$_{0.15}$, \textit{i.e.} $\textrm{Cu}_x(\textrm{Ni}_{3/4}\textrm{Al}_{1/4})_{1-x}$ for $x=0.4$}
    \label{fig:x_0.40}
\end{figure}

\begin{figure}[hp]
    \centering
    \includegraphics[width=\linewidth]{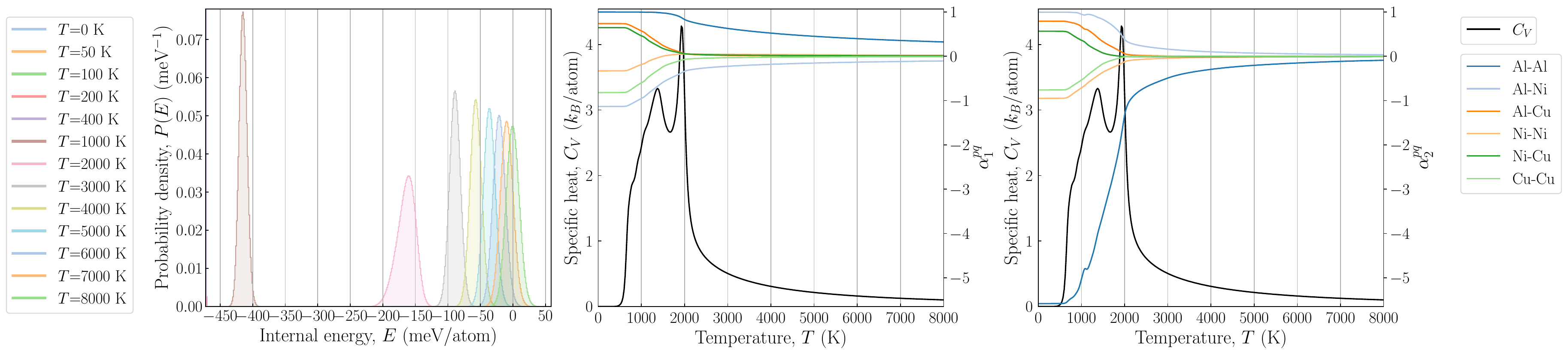}
    \caption{Obtained specific heat ($C_V$) and Warren-Cowley ASRO parameters ($\alpha^{n}_{pq}$) as a function of temperature for Cu$_{0.45}$Ni$_{0.4125}$Al$_{0.1375}$, \textit{i.e.} $\textrm{Cu}_x(\textrm{Ni}_{3/4}\textrm{Al}_{1/4})_{1-x}$ for $x=0.45$}
    \label{fig:x_0.45}
\end{figure}

\begin{figure}[hp]
    \centering
    \includegraphics[width=\linewidth]{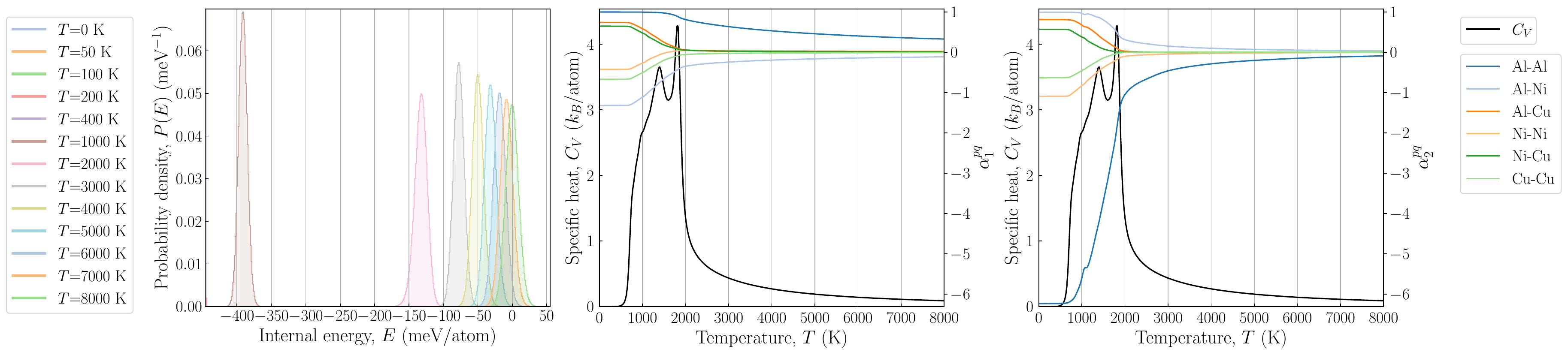}
    \caption{Obtained specific heat ($C_V$) and Warren-Cowley ASRO parameters ($\alpha^{n}_{pq}$) as a function of temperature for Cu$_{0.5}$Ni$_{0.375}$Al$_{0.125}$, \textit{i.e.} $\textrm{Cu}_x(\textrm{Ni}_{3/4}\textrm{Al}_{1/4})_{1-x}$ for $x=0.5$}
    \label{fig:x_0.50}
\end{figure}

\begin{figure}[hp]
    \centering
    \includegraphics[width=\linewidth]{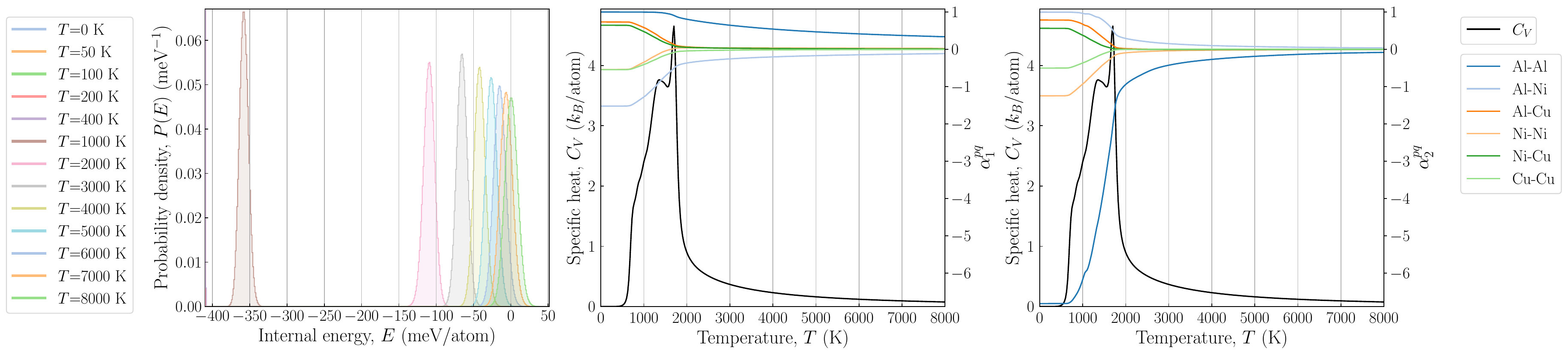}
    \caption{Obtained specific heat ($C_V$) and Warren-Cowley ASRO parameters ($\alpha^{n}_{pq}$) as a function of temperature for Cu$_{0.55}$Ni$_{0.3375}$Al$_{0.1125}$, \textit{i.e.} $\textrm{Cu}_x(\textrm{Ni}_{3/4}\textrm{Al}_{1/4})_{1-x}$ for $x=0.55$}
    \label{fig:x_0.55}
\end{figure}

\begin{figure}[hp]
    \centering
    \includegraphics[width=\linewidth]{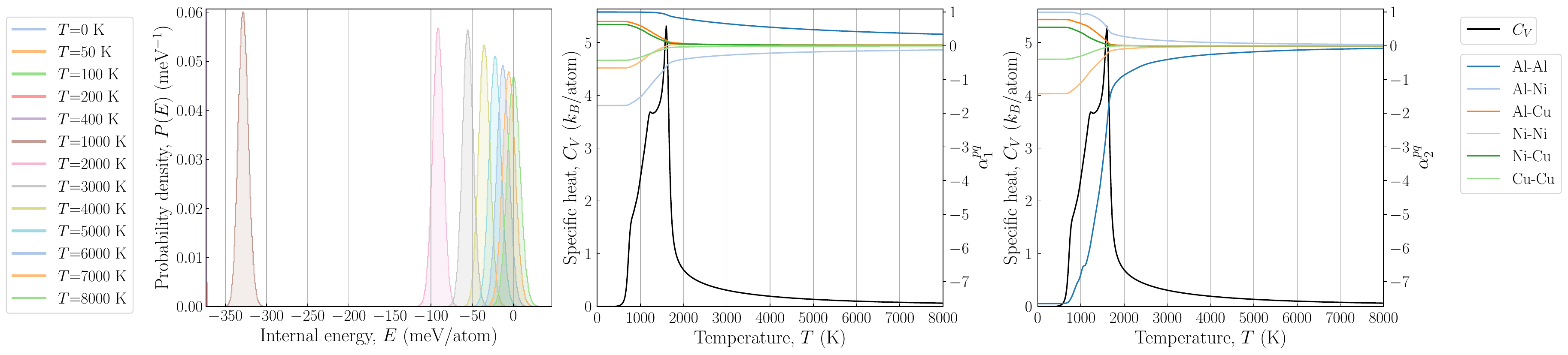}
    \caption{Obtained specific heat ($C_V$) and Warren-Cowley ASRO parameters ($\alpha^{n}_{pq}$) as a function of temperature for Cu$_{0.6}$Ni$_{0.3}$Al$_{0.1}$, \textit{i.e.} $\textrm{Cu}_x(\textrm{Ni}_{3/4}\textrm{Al}_{1/4})_{1-x}$ for $x=0.6$}
    \label{fig:x_0.60}
\end{figure}

\begin{figure}[hp]
    \centering
    \includegraphics[width=\linewidth]{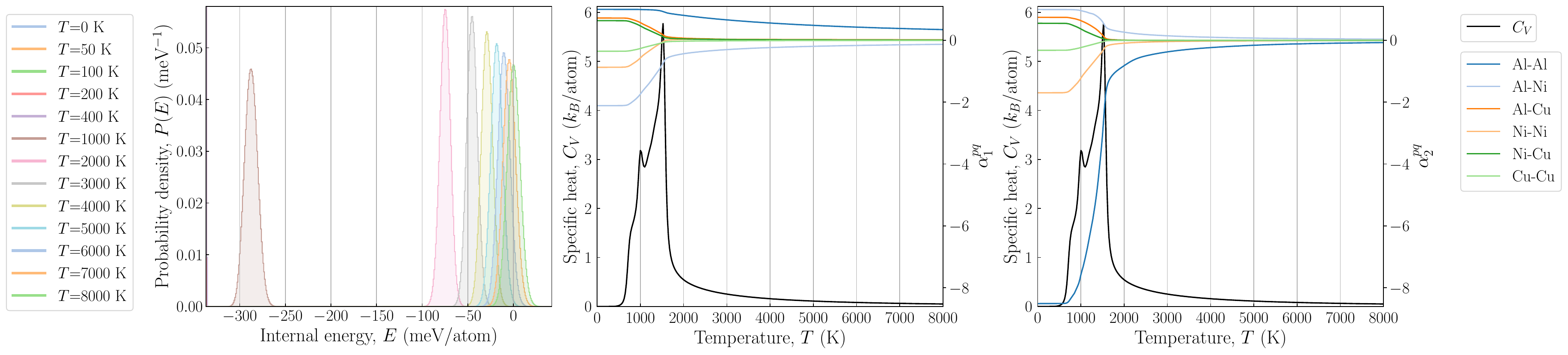}
    \caption{Obtained specific heat ($C_V$) and Warren-Cowley ASRO parameters ($\alpha^{n}_{pq}$) as a function of temperature for Cu$_{0.65}$Ni$_{0.2625}$Al$_{0.0875}$, \textit{i.e.} $\textrm{Cu}_x(\textrm{Ni}_{3/4}\textrm{Al}_{1/4})_{1-x}$ for $x=0.65$}
    \label{fig:x_0.65}
\end{figure}

\begin{figure}[hp]
    \centering
    \includegraphics[width=\linewidth]{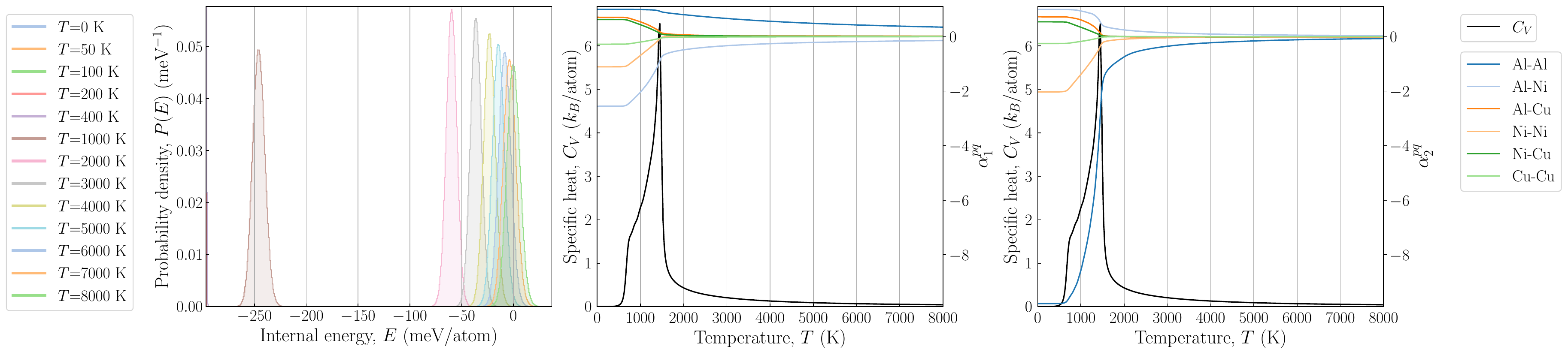}
    \caption{Obtained specific heat ($C_V$) and Warren-Cowley ASRO parameters ($\alpha^{n}_{pq}$) as a function of temperature for Cu$_{0.7}$Ni$_{0.225}$Al$_{0.075}$, \textit{i.e.} $\textrm{Cu}_x(\textrm{Ni}_{3/4}\textrm{Al}_{1/4})_{1-x}$ for $x=0.7$}
    \label{fig:x_0.70}
\end{figure}

\begin{figure}[hp]
    \centering
    \includegraphics[width=\linewidth]{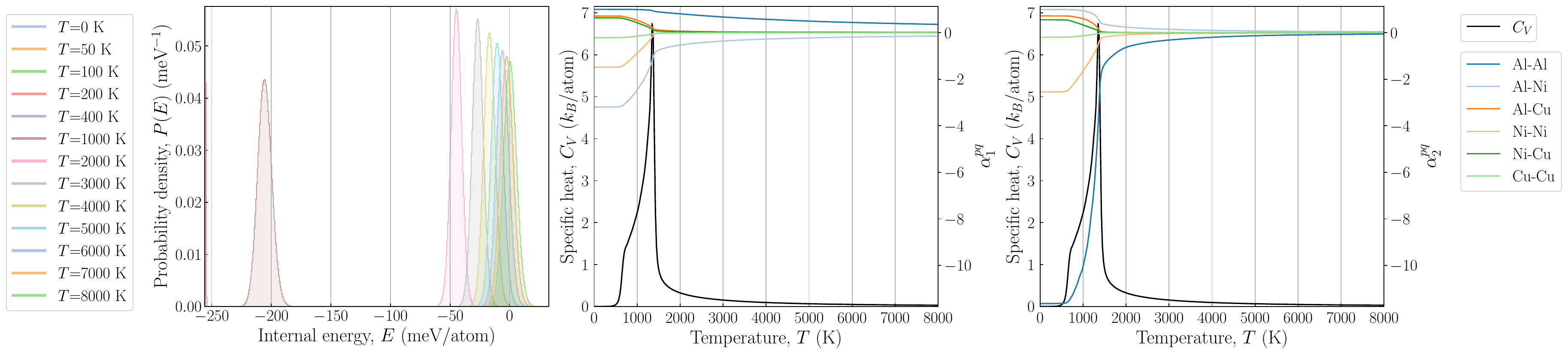}
    \caption{Obtained specific heat ($C_V$) and Warren-Cowley ASRO parameters ($\alpha^{n}_{pq}$) as a function of temperature for Cu$_{0.75}$Ni$_{0.1875}$Al$_{0.0625}$, \textit{i.e.} $\textrm{Cu}_x(\textrm{Ni}_{3/4}\textrm{Al}_{1/4})_{1-x}$ for $x=0.75$}
    \label{fig:x_0.75}
\end{figure}

\begin{figure}[hp]
    \centering
    \includegraphics[width=\linewidth]{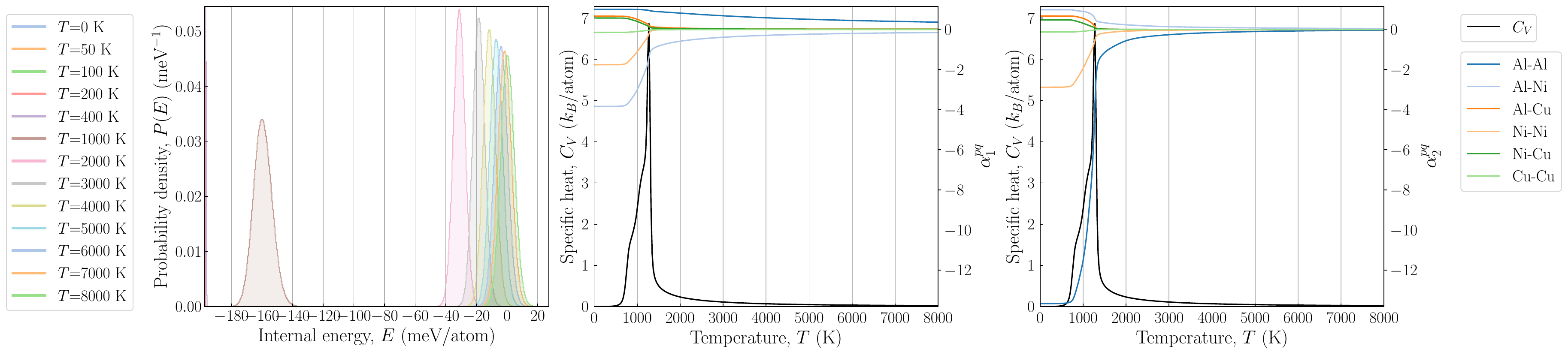}
    \caption{Obtained specific heat ($C_V$) and Warren-Cowley ASRO parameters ($\alpha^{n}_{pq}$) as a function of temperature for Cu$_{0.8}$Ni$_{0.15}$Al$_{0.05}$, \textit{i.e.} $\textrm{Cu}_x(\textrm{Ni}_{3/4}\textrm{Al}_{1/4})_{1-x}$ for $x=0.8$}
    \label{fig:x_0.80}
\end{figure}

\begin{figure}[hp]
    \centering
    \includegraphics[width=\linewidth]{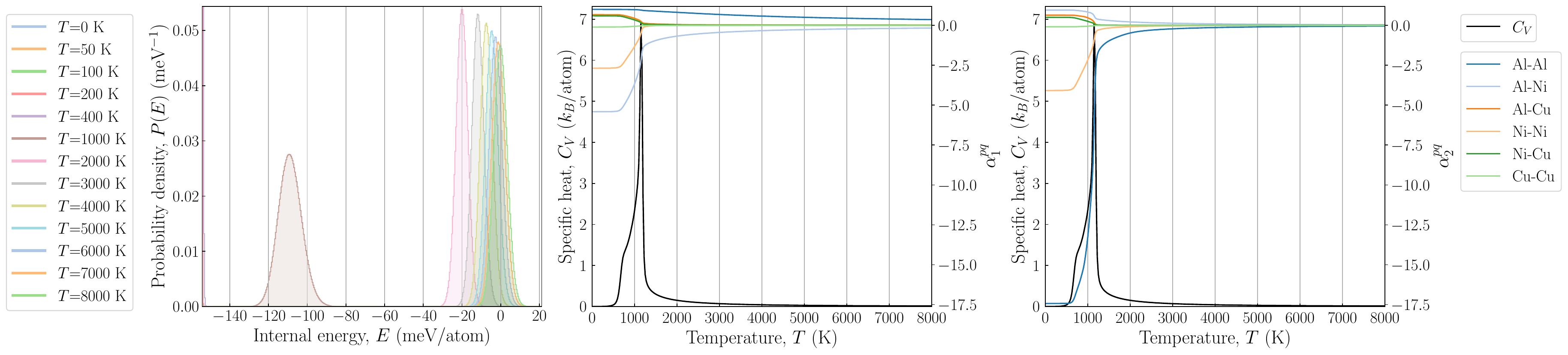}
    \caption{Obtained specific heat ($C_V$) and Warren-Cowley ASRO parameters ($\alpha^{n}_{pq}$) as a function of temperature for Cu$_{0.85}$Ni$_{0.1125}$Al$_{0.0375}$, \textit{i.e.} $\textrm{Cu}_x(\textrm{Ni}_{3/4}\textrm{Al}_{1/4})_{1-x}$ for $x=0.85$}
    \label{fig:x_0.85}
\end{figure}

\begin{figure}[hp]
    \centering
    \includegraphics[width=\linewidth]{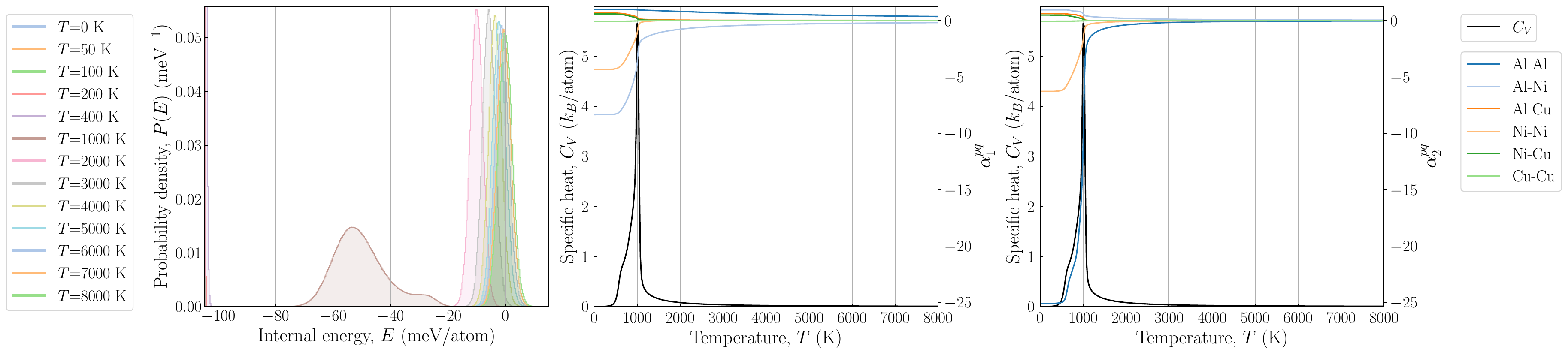}
    \caption{Obtained specific heat ($C_V$) and Warren-Cowley ASRO parameters ($\alpha^{n}_{pq}$) as a function of temperature for Cu$_{0.9}$Ni$_{0.075}$Al$_{0.025}$, \textit{i.e.} $\textrm{Cu}_x(\textrm{Ni}_{3/4}\textrm{Al}_{1/4})_{1-x}$ for $x=0.9$}
    \label{fig:x_0.90}
\end{figure}

\begin{figure}[hp]
    \centering
    \includegraphics[width=\linewidth]{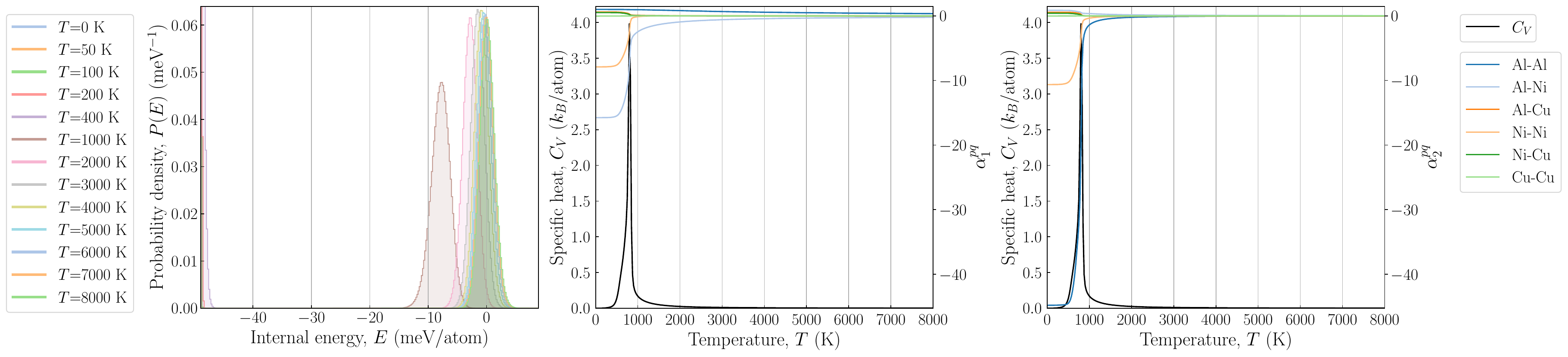}
    \caption{Obtained specific heat ($C_V$) and Warren-Cowley ASRO parameters ($\alpha^{n}_{pq}$) as a function of temperature for Cu$_{0.95}$Ni$_{0.0375}$Al$_{0.0125}$, \textit{i.e.} $\textrm{Cu}_x(\textrm{Ni}_{3/4}\textrm{Al}_{1/4})_{1-x}$ for $x=0.95$}
    \label{fig:x_0.95}
\end{figure}

Figs.~\ref{fig:x_0.00} through \ref{fig:x_0.95} present combined plots of the temperature dependent specific heat and Warren--Cowley ASRO parameters for $\textrm{Cu}_x(\textrm{Ni}_{3/4}\textrm{Al}_{1/4})_{1-x}$ for $x=0$ through to $x=0.95$. All results presented here are obtained using the Wang--Landau sampling algorithm~\cite{wang_efficient_2001, wang_determining_2001} as implemented in the \textsc{Brawl} package~\cite{naguszewski_brawl_2025}, with full computational details outlined in the main text and in the open-access dataset associated with this work~\cite{dataset}.

\newpage

